%% file: Hyndman-Wenger-1410.7453v2-nc.tex
\documentclass[paper=letter, fontsize=12pt]{article} 

\usepackage{url}
\usepackage{lifecon}
\usepackage{prettyref}
\usepackage{ctable}
\usepackage{threeparttable}
\usepackage{booktabs}
\usepackage{subfig}
\usepackage{todonotes}
\usepackage{appendix}

\newcommand{\al}{\alpha}
\newcommand{\faira}{\alpha^{\star}}
\DeclareMathOperator{\Q}{\mathbb{Q}}
\newcommand{\bara}{\bar{\alpha}}
\newcommand{\barr}{\bar{r}}

\newcommand{\bartauzero}{\bar{\tau}_{0}}

\newcommand{\mb}[1]{\mathbb{#1}}
\newcommand{\mc}[1]{\mathcal{#1}}
\newcommand{\CC}{C\nolinebreak[4]\hspace{-.05em}\raisebox{.4ex}{\tiny\bf ++}}

\newrefformat{tab}{Table\,\,\ref{#1}}
\newrefformat{fig}{Figure\,\,\ref{#1}}
\newrefformat{eq}{(\ref{#1})}
\newrefformat{thm}{Theorem\,\,\ref{#1}}
\newrefformat{def}{Definition\,\,\ref{#1}}
\newrefformat{rem}{Remark\,\,\ref{#1}}
\newrefformat{lem}{Lemma\,\,\ref{#1}}
\newrefformat{pro}{Proposition\,\,\ref{#1}}
\newrefformat{ass}{Assumption\,\,\ref{#1}}
\newrefformat{exa}{Example\,\,\ref{#1}}
\newrefformat{sec}{Section\,\,\ref{#1}}
\newrefformat{sub}{Subsection\,\,\ref{#1}}
\newrefformat{cha}{Chapter\,\,\ref{#1}}

\usepackage{amssymb}
\usepackage{amsthm}
\RequirePackage{amsmath,amsfonts}
\RequirePackage[numbers]{natbib}
\RequirePackage[colorlinks,citecolor=blue,urlcolor=blue]{hyperref}

\newcounter{thm}
 \newcounter{ex}
 \newcounter{re}
 \newtheorem{Theorem}[thm]{Theorem}
 \newtheorem{Lemma}[thm]{Lemma}
 
 \newtheorem{Proposition}[thm]{Proposition}
\newtheorem{Assumption}[thm]{Assumption}

 \newtheorem{Example}[ex]{Example}
 
 \newtheorem{Remark}[re]{Remark}
 \newtheorem{Definition}[thm]{Definition}

\newlength{\oldparindent}
\RequirePackage[OT1]{fontenc}
\usepackage{keyval}
\usepackage{calc}
\usepackage{fancyhdr}
\begin{document}

\rhead{\textit{\today}}
\lhead{\textit{ C.\ Hyndman \& M.\ Wenger }}
\chead{\textit{\quad \quad GMWB Riders in a Binomial Framework}}

\title{GMWB Riders in a Binomial Framework - Pricing, Hedging, and Diversification of Mortality Risk\footnote{This paper combines a previous version  titled ``Pricing and Hedging GMWB Riders in a Binomial Framework'' (\textit{arXiv:1410.7453v1}) and the working paper titled ``Diversification of mortality risk in GMWB rider pricing and hedging''}}

\author{Cody HYNDMAN\footnote{Corresponding Author: email: cody.hyndman@concordia.ca}\ \footnote{ 
Department of Mathematics and Statistics, 
Concordia University, 
1455 Boulevard de Maisonneuve Ouest,
Montr\'eal, Qu\'ebec,
Canada H3G 1M8.
} \ and \  Menachem~WENGER\footnote{The Guardian Life Insurance Company of America, New York, NY}\ \footnote{The views and opinions expressed in this paper are those of the individual author(s) and do not necessarily reflect the views of The Guardian Life Insurance Company of America.}
\ 
}

\date{\today}

\maketitle

\abstract{
We construct a binomial model for a guaranteed minimum withdrawal benefit
(GMWB) rider to a variable annuity (VA) under optimal policyholder behaviour.  
The binomial model results in explicitly formulated perfect hedging strategies
funded using only periodic fee income.
We consider the separate perspectives of the insurer and policyholder and 
introduce a unifying relationship.
Decompositions of the VA and GMWB contract into term-certain payments and options
representing the guarantee and early surrender features 
are extended to the binomial framework.
We incorporate an approximation algorithm for Asian options that significantly improves efficiency of
the binomial model while retaining accuracy. Several numerical examples
are provided which illustrate both the accuracy and the tractability
of the binomial model. We extend the binomial model to include policy holder mortality and death benefits.  Pricing, hedging, and the decompositions of the contract are extended to incorporate mortality risk.  We prove limiting results for the hedging strategies and demonstrate mortality risk diversification.  Numerical examples are provided which illustrate the effectiveness of hedging and the diversification of mortality risk under capacity constraints with finite pools.
}

\vspace{5mm}

\noindent
\textbf{Keywords:}
variable annuity; GMWB; optimal stopping; hedging; binomial models; mortality

\vspace{5mm}
\noindent
\textbf{Mathematics Subject Classification (2010):}
Primary: 91G20; 91G60; Secondary: 91B30; 60G40

\vspace{5mm}
\noindent
\textbf{JEL Classification:} G22, G12, G13, C61, C63

\vfill
\pagebreak

\renewcommand{\baselinestretch}{1.5}

\input{Hyndman-Wenger-Content-nc}

\vspace{1em}
\noindent\textbf{Acknowledgements}
This research was supported by the Natural Sciences and Engineering Research Council (NSERC) of Canada and the Fonds de recherche du Qu\'ebec - Nature et technologies (FQRNT).  

\appendixtitleon
\begin{appendices}

  \input{Hyndman-Wenger-Appendix-nc}

\end{appendices}

\vspace{1em}
\noindent
\bibliographystyle{abbrvnat}
\bibliography{wmendy-RefList1}
\end{document}

%% file: Hyndman-Wenger-Content-nc.tex
\section{\label{sec:Introduction}Introduction}

The variable annuity (VA) with guaranteed minimum withdrawal benefit (GMWB) rider
was introduced in 2002.  These contracts allow for an accumulation period
where an initial premium deposited with the insurer is invested in a  portfolio
of funds selected by the policyholder.  The account value (AV) benefits from
gains made by the portfolio and a periodic fee is deducted by the insurer. 
The policy holder can take periodic withdrawals from the AV, 
up to certain limits, and cumulative withdrawals are guaranteed to return 
the initial premium over the term of the contract.  The contract may be 
surrendered early, enabling the policyholder to benefit from strong portfolio performance, 
subject to contingent deferred sales charges (CDSC).  At the end of the term, provided the
contract has not already been surrendered, the contract may be annuitized for either
a fixed term or the remaining life of the policyholder. A large literature on the modeling 
and pricing of these contracts, as well as other forms of guarantees, has emerged since their introduction
to the marketplace.  A brief overview of the history of GMWB and similar products as well as the 
various modeling and pricing approaches can be found in \citet{Hyndman-Wenger-Decomp}.

Around the time of the financial crisis in 2008 reinsurers stopped offering
coverage altogether on GMWB and related guaranteed lifetime withdrawal
benefit (GLWB) riders at which point the importance
of internal dynamic hedging programs rose rapidly.  With this in mind, we consider the problems of
pricing and hedging the GMWB product in a discrete time framework consistent with the no-arbitrage principle
from financial economics.  We propose a binomial asset pricing model for GMWBs assuming optimal policyholder behaviour
and construct explicit hedging strategies. An overview of other approaches to policyholder behaviour can be found in \citet{Kling2014}, \citet{doi:10.1080/14697688.2013.825922}, and the references therein.

The binomial model has several advantages which we believe justify its use in theory and practice.
It is significantly simpler to obtain numerical results using the binomial model than many of the
approaches which have previously been applied.
Under an appropriate parameterization the binomial model converges to the \citet{RefWorks:99} model, 
which has been used as the basis for modeling these contracts by a number of authors,
and yields good approximations for more complex financial options which lack analytic solutions in the corresponding
 continuous time pricing models. Through dynamic programming and backward induction algorithms, binomial
pricing models can easily be implemented. Further, the binomial model can be calibrated to a volatility surface.

In contrast to Monte-Carlo simulation methods, the binomial
approach is well-suited for American-style options with early exercise capability.
More importantly an explicit exact hedging strategy can be formulated and implemented.
Although binomial methods can be seen as a special case of finite difference methods 
there are fundamental differences between the two general methods and a thorough comparison of binomial and finite difference methods is provided in \citet{RefWorks:100}.

Binomial models are ideally suited for non path-dependent products.
In such a setting, aside from enabling a simple theoretical framework,
it is computationally efficient to obtain reliable numerical results.
The GMWB product is path-dependent and we discuss the implications of this 
and address them by employing an approximation technique.  Although in theory the results should converge
to those of the continuous withdrawal model where the investment fund is log-normally
distributed; due to the non-recombining nature of the account value
the suggested method is found to be numerically expensive. We substantially
improve the numerical efficiency without sacrificing significant accuracy
of results by adopting an approximation method based on \citet{RefWorks:84}. 

A binomial valuation approach has previously been considered by \citet{RefWorks:116}
to price equity-linked life insurance with recurring premiums in the
presence of early surrenders. Although the underlying methodology
is similar, we deal with the unique features and challenges of modeling
GMWB riders for variable annuities. In addition to surrender and mortality,
both elements considered by \citet{RefWorks:116}, we have an endogenously
determined trigger date. The nature of the fees and withdrawals further
differentiate our work. Whereas \citet{RefWorks:116} deals exclusively
with pricing, we pay equal attention to the hedging constructions
in a binomial model, which is facilitated by the consideration of
the perspectives of both the insurer and insured. By focusing on
a single product we have the liberty to consider a top-down approach
which provides more insight than generic formulations of backward
induction schemes.

The remainder of this paper is organized as follows.  In \prettyref{sec:Bin wo mort}
we present the binomial asset pricing model for variable annuities with GMWBs riders in a  restricted model which accounts for equity risk only.  We extend the model in \prettyref{sec:Binlapses} to allow
for surrenders - that is, we incorporate behaviour risk.  In \prettyref{sec:Comput} we discuss computational considerations in the implementation of the binomial model and we present a binomial approximation algorithm
designed to improve numerical efficiency. In \prettyref{sec:Num-Res-No-Mort} numerical results using the binomial model are obtained and compared with results from the literature.  In \prettyref{sec:bin mort} we extend the binomial model to include mortality risk and death benefits as well as proving mortality diversification results and considering numerical experiments reflecting capacity constraints and finite pools of policy holders. Section~\ref{sec:Conc} concludes and an Appendix contains technical results and proofs.

\section{GMWBs in a Binomial Asset Pricing Model}
\label{sec:Bin wo mort}

In this section we define and construct a binomial asset pricing model for the variable annuity with GMWB rider.
We first introduce the product specifications and notation following \citet{RefWorks:33} and 
\citet{RefWorks:63} for the binomial model and \citet{Hyndman-Wenger-Decomp} for the variable annuity with GMWB rider. 

\subsection{Contract Specifications and Model Framework}

At time $t=0$, a policy, consisting of an underlying variable annuity (VA) contract and a GMWB rider, is issued to
a policyholder of age $x$ and an initial premium $P$ is received.
We assume no subsequent premiums are paid after time zero. The premium is invested into a fund
which tracks the price of  a risky asset $S=\{S_{t}: t\geq 0 \}$ with no basis risk. The
rider fee rate $\al$ is periodically discounted from the account value $W=\{W_{t};t \geq 0 \}$ 
as long as the contract is in force and the account value is positive. 

The GMWB rider contract specifies a guaranteed maximal withdrawal rate $g$ so that $G:=gP$ can be withdrawn annually until the initial premium is recovered 
regardless of the evolution of $\{W_{t}\}$.  If the account value falls to zero the policyholder continues 
to make withdrawals at rate $G$ until the initial premium has been recovered. 
Policyholders may withdraw any amount from the account value not exceeding 
the remaining account value.  However, if annual withdrawals
exceed $G$ while the account value is still positive then a surrender
charge is applied to the withdrawals and a reset feature may reduce
the guarantee value, i.e. the remaining portion of the initial premium
not yet recovered. Policyholders also have the option of surrendering early  
and receiving the account value less a surrender charge. The terminology of lapses and surrenders are used interchangeably. Any guarantee value is forfeited by surrendering. 

Assuming a static withdrawal strategy where $G$ is withdrawn annually we set the maturity $T:=1/g$ since
the sum of all withdrawals at $T$ is $P$. At time $T$ the
rider guarantee is worthless and the policyholder receives a terminal
payoff of the remaining account value if it is positive. 
This assumption translates over to a real-world trend of no annuitizations and is justified since a high proportion
of VAs are not ultimately annuitized.

Consider a financial market consisting of one risky
asset $S$ and a riskless money market account growing at a constant continuously compounded risk-free interest rate $r$. Let $n$
be the number of time-steps per year, $N=T\times n$ the total
number of time-steps modelled, and $\delta t=1/n$ the length of each
time-step. For $i\in\mc{I}_{N}^{+}:=\{1,\dots,N-1,N\},$ write
$S_{i}$ for the asset value at time $i\delta t$.
We assume that the insurer can borrow and lend at rate $r$.
Given $S_{i-1}$, the asset value $S_{i}$ takes one of two values: $S_{i-1}u$ or $S_{i-1}d$, 
where $u$ ($d$) represents an up-movement (down-movement) in the asset value.
To rule out arbitrage opportunities and the trivial case of no randomness, $u$
and $d$ must satisfy 

\begin{equation}
0<d<e^{r\delta t}<u\label{eq:r constraint}
\end{equation}

\noindent as in \citet{RefWorks:33}.

Consider a sequence of $N$ coin tosses. Let $\Omega:=\{H,T\}^{N}$
and $\mc{F}:=2^{\Omega}$. Denote a sample point of $\Omega$
by $\bar{\omega}_{N}:=\omega_{1}\dots\omega_{N}:=(\omega_{1},\dots,\omega_{N})$.
Consider the stochastic process $\xi=(\xi_{i})_{1\leq i\leq N}$,
where $\xi_{i}:\Omega\mapsto\{u,d\}$ is

$$
\xi_{i}(\bar{\omega}_{N})=\begin{cases}
u & \text{if }\omega_{i}=H,\\
d & \text{if }\omega_{i}=T.\end{cases}
$$

\noindent Then for any fixed $\bar{\omega}_{N}$, $\xi_{i}(\bar{\omega}_{N})$
maps $i$ to the growth factor of $S$ in period $i$. The natural
filtration is $\mc{F}_{i}=\sigma(\xi_{j};\ j\leq i)$. 

The $\mb{F}$-adapted process $\{S_{i}\}$ can be expressed as $S_{i}=S_{0}\times\prod_{j=1}^{i}\xi_{j}$
where $S_{0}$ is the initial value of the risky asset. 
We write $\bar{\omega}_{i}=\omega_{1}\dots\omega_{i}$
to refer to the specific path evolution up to time $i$. For any $j\leq i$,
we write

\[
\xi_{j}(\bar{\omega}_{i})=\begin{cases}
u & \text{if }\omega_{j}=H,\\
d & \text{if }\omega_{j}=T.\end{cases}
\]

\noindent 
Finally, we replace $H$ and $T$ with $u$ and $d$ respectively when defining
$\Omega$, therefore the sample path $\bar{\omega}_{N}$ refers directly
to the evolution of the underlying asset $S$ where each $\omega_{j}\in\{u,d\}$.
Then for any $\bar{\omega}_{i}$, 

\begin{equation}
S_{i}=S_{0}\prod_{j=1}^{i}\xi_{j}(\bar{\omega}_{i})=S_{0}\prod_{j=1}^{i}\omega_{j}=S_{0}u^{\{\text{\# of }u\text{ in }\bar{\omega}_{i}\}}d^{\{\text{\# of }d\text{ in }\bar{\omega}_{i}\}}.\label{eq:SiS0Notation}
\end{equation}

\noindent
Beginning with $S_{0}=P$, the binomial tree for $\{S_{i}\}$ is constructed
forward in time. For $i\in\mc{I}_{N}^{+}$, set

$$
S_{i}=\xi_{i}S_{i-1}.
$$

The unique risk-neutral measure
$\Q$ is defined on 
$(\Omega,\mc{F}_{N})$ by 

\[
 \Q(A):=\sum_{\bar{\omega}_{N}\in A}p^{\{\#\text{ of }u\text{ in }\bar{\omega}_{N}\}}q^{N-\{\#\text{ of }u\text{ in }\bar{\omega}_{N}\}}
\] 

\noindent
for any set $A\in\mc{F}_{N}$ where 

\begin{equation}
p:=\frac{e^{r\delta t}-d}{u-d},\label{eq:RNP}
\end{equation}

\noindent
is the risk-neutral probability of observing
an $H$ at any particular coin toss (observing a
$u$ at any particular time step), $q:=1-p$, and $p>0$. 
The constructed probability space is $(\Omega,\mc{F}_{N},\mb{F}=\{\mc{F}_{i}\}_{0\leq i\leq N},\Q)$. 
Note that $p\in(0,1)$ by \prettyref{eq:r constraint} and there are no
$(\Q,\mc{F}_{N})$-negligible sets and so all results
hold for all $\omega\in\Omega$.

We follow the Cox, Ross, and Rubinstein (CRR) parametrization and set 
$u=\exp{(\sigma\sqrt{\delta t})}$ and $d=\exp{(-\sigma\sqrt{\delta t})}$, where $\sigma$ is the variance of the continuously compounded rate of return of $S$.  The CRR parametrization leads to the following result.
\begin{Proposition}
Suppose $S_{t}$ follows the dynamics given by 

\begin{equation}
dS_{t} = r S_{t} dt + \sigma S_{t} dB_{t}
\label{eq:StP} \end{equation}

\noindent
where $B_{t}$ is a standard Brownian motion.
Consider the binomial model for $S_{i}^{n}$ with $n$ time-steps per year under the CRR parametrization. Then for all $t\in[0,T]$, as $n\to\infty$, $S_{nt}^{n}$ converges
in distribution to $S_{t}$, where $nt$ is an integer.
\end{Proposition}
\begin{proof}
See \citet{RefWorks:46} or \citet[Exercise 3.8]{RefWorks:41}. 
\end{proof}
Note that the real world probability measure $\mb{P}$ is defined similarly but with
 $\tilde{p}=\frac{1}{2}+\frac{1}{2}\frac{\hat{\mu}}{\hat{\sigma}}\sqrt{\delta t}$  where $\hat{\mu}$ and $\hat{\sigma}$ are the respective empirical mean and variance of the continuously compounded rate of return of $S$. Under this parametrization, the mean and variance under the binomial model converge to the empirical values in the limit (see \citet{RefWorks:46}).

We specify the underlying assumptions on the variable annuity with GMWB rider that are employed throughout this section.
\begin{Assumption}
\label{ass:bin no lapse}Early surrenders
are not allowed. Under the static withdrawal strategy the policyholder
receives $G=gP\delta t$ each time period. We set $T:=1/g$. 
At the end of each period the pro-rated rider fee is first deducted
and then the periodic withdrawal is subtracted. We restrict $r>0$ and denote $\barr:=r\delta t$ and $\bar{\alpha}:=\alpha\delta t$.
\end{Assumption}

\begin{Remark}
We assume $T$ is an integer. Otherwise, the
results can be adapted to incorporate the final fractional period.
Set $N=\lfloor T\cdot n\rfloor+1$ and the final period has time length
of $T-\left(\frac{N-1}{n}\right)$years. All the parameters need to be scaled for the terminal period
to reflect the shortened duration. %
\end{Remark}

Next, we define another binomial tree for the account value $W$ which contains two values at each node. The first component, denoted $W_{i^{-}}$, is the account value after adjusting for market movements but before fees are deducted or withdrawals are made.  The second component, denoted $W_{i}$, is the account value after adjusting for fees and withdrawals. 
We have
 
\begin{align*}
&W_{0} =P,\\
&W_{i^{-}} =\frac{S_{i}}{S_{i-1}}W_{i-1}=\xi_{i}W_{i-1},\\
&W_{i} =\max\left\{e^{-\bara} W_{i^{-}}-G,0\right\} ,\end{align*}

\noindent
for $i\in\mc{I}_{N}^{+}$. Although the tree for the underlying
asset $\{S_{i}\}$ is recombining, the tree for the account value
$\{W_{i}\}$ is non-recombining. For any $i$ there are $i+1$ nodes
for $S_{i}$ but $2^{i}$ nodes for $W_{i}$ on the respective trees.
The subtraction of the periodic withdrawals imposes a path dependency
on the model.

\subsection{Valuation perspectives and decompositions}

There are two separate perspectives for valuing the variable annuity with GMWB rider.  The first, corresponding to the policyholder,
treats the variable annuity and GMWB rider together and values the total payments received over the life of the contract.  The second, corresponding to the insurer, considers the embedded optionality of the GMWB rider separately to price and hedge the additional risk.
This approach was used in \citet{RefWorks:32} and \citet{Hyndman-Wenger-Decomp} in a continuous time setting.  In the  discrete-time binomial model we obtain similar theoretical results as in the continuous-time setting of \citet{Hyndman-Wenger-Decomp}. However, in the discrete-time  binomial model we also provide an explicit computational framework for pricing and hedging.

\subsubsection{Policyholder valuation perspective}

Denote by $V_{n}$ the value to the policyholder at time $n$ of the remaining payments to be received from the complete contract (VA plus GMWB rider).  By the risk neutral pricing formula we obtain the following backward in time recursive relationship

\begin{align}
V_{N} & =W_{N},\nonumber \\
V_{i} & =E_{\Q}\left[\sum_{m=i+1}^{N}Ge^{-\barr(m-i)}+e^{-\barr(N-i)}W_{N}|\mc{F}_{i}\right]\nonumber \\
 & =Ga_{\lcroof{N-i}}+e^{-\barr(N-i)}E_{\Q}[W_{N}|\mc{F}_{i}]\label{eq:V_0Discrete}\end{align}

\noindent
for $i\in\mc{I}_{N-1}$, with $\mc{I}_{N-1}:=[0,1,\dots,N-1]$
and $a_{\lcroof{m}}=(1-e^{-\barr m})/(e^{\barr}-1)$. For $i=0$
this reduces to 

\begin{equation}
   V_{0}=Ga_{\lcroof{N}}+e^{-\barr N}E_{\Q}[W_{N}]. \label{eq:V0Bin}
\end{equation}

\noindent
Note that equations~(\ref{eq:V_0Discrete}) and (\ref{eq:V0Bin}) are 
the discrete-time analogues to the policyholder's valuation given, respectively, by equations (7) and (6) of \citet{Hyndman-Wenger-Decomp}.  We write $V_{0}:=V_{0}(P,\alpha,g)$ when we wish to emphasize the dependence on the contract parameters of the value to the policyholder.  

The process $\{V_{i}\}$ represents the value of the combined annuity
plus GMWB rider contract at each time point just after the deduction
of fees and withdrawals. By the Markov property we have $V_{i}=v(i,W_{i})$,
 where $v:\mc{I}_{N}\times\mb{R}_{+}\mapsto\mb{R}_{+}$
is

\begin{equation}
v(i,x)=\begin{cases}
x & i=N,\\
{}[G+pv(i+1,w(xu))+qv(i+1,w(xd))]e^{-\barr} & i<N,\end{cases}\label{eq:V0BinNolapse}\end{equation}

\noindent
and $w:\mb{R}_{+}\mapsto\mb{R}_{+}$ is given by

\begin{equation}
w(x)=\max\{xe^{-\bara}-G,0\}.\label{eq:w(x)}\end{equation}

\noindent
Note that $\{e^{-\barr i}V_{i}+Ga_{\lcroof{i}}\}_{0\leq i\leq N}$
is a $(\Q,\mb{F})$ martingale for all $\alpha$.

As in \citet{Hyndman-Wenger-Decomp} we define the fair fee rate as follows.
\begin{Definition}\label{defn:fair-alpha}
A \textit{fair fee rate} is a rate $\faira\geq 0$ such that

\begin{equation}
V_{0}(P,\faira,g) = P \label{def:fairfeerate}.
\end{equation}

\noindent
\end{Definition}
There is no closed form solution for $\faira$. However, as in \citet{Hyndman-Wenger-Decomp}, we are able to prove the existence
and uniqueness of the fair fee rate by showing that the value $V_{0}$ is continuous and monotone as a function of $\alpha$.  However, in a finite probability space $\Q(W_{N}>0)=0$ for sufficiently large $\alpha$. Consequently strict monotonicity holds only on a bounded interval. 
\begin{Lemma}
\label{lem:MonotBin}For all fixed $(i,x)\in\mc{I}_{N-1}\times\mb{R}_{++}$,
the contract value function $v(i,x)$, defined by \prettyref{eq:V0BinNolapse},
as a function of $\alpha$ is continuous for $\alpha\geq0$ and strictly
decreasing on $[0,b^{x,i})$ where

\[
b^{x,i}:=\min\{\alpha\geq0:W_{N}^{x,i}=0\text{ a.s.}\}<\infty.\] 

\noindent
Further, if $(i,x)$ satisfies 

\begin{equation}
x>G\sum_{j=1}^{N-i}d^{j}\label{eq:x>Gsum...}\end{equation}

\noindent
then $b^{x,i}>0$, otherwise $b^{x,i}=0$. For $\alpha\geq b^{x,i}$,
$v(x,i)=Ga_{\lcroof{N-i}}$. \end{Lemma}
\begin{proof}
See Appendix~\ref{Proofs}.
\end{proof}

In particular, equation~\prettyref{eq:x>Gsum...} holds for $(i,x)=(0,P)$
since $G=P/N$ and $d<1$. The existence and uniqueness of $\faira$
is discussed in the next theorem.
\begin{Theorem}
\label{thm:existuniquealpha}Under \prettyref{ass:bin no lapse} there
exists a unique $\faira\in[0,b^{P,0})$ such that $V_{0}(P,\faira,g)=P$.
\end{Theorem}
\begin{proof}
See Appendix~\ref{Proofs}.
\end{proof}

\begin{Remark}
For $r=0$ we have $V_{0}(P,\alpha,g)=P$ for all $\alpha\geq b^{P,0}$.
Thus $r>0$ is a necessary condition to ensure uniqueness of $\faira$.
\end{Remark}

From Lemma~\ref{lem:MonotBin} we may iteratively solve for the fair fee using the bisection method provided we have a method for calculating the value $V_{0}$ as a function of $\alpha$.  We shall discuss the technical details of this process and computational challenges after consideration of the insurer's valuation perspective, hedging, and the extension of the model to include lapses. %

\subsubsection{\label{sub:Insurer-Perspective}Insurer valuation perspective}

The insurer may consider the guarantees embedded in the variable annuity contract as separate products.  From this point of view it is necessary to consider the time at which the account value hits zero and subsequent payments to the policyholder are drawn from the guarantee.  Define the discrete-time analogue of the trigger time of the continuous model considered by \citet{RefWorks:7} as follows.
\begin{Definition}
In the binomial model, the trigger time $\tau$ is defined as the
stopping time

\[
\tau(\omega_{1}\dots\omega_{N}):=\inf\{i\geq1;W_{i}(\omega_{1}\dots\omega_{i})=0\},\] 

\noindent
where $\inf(\emptyset)=\infty.$ For any fixed sequence $\bar{\omega}_{i}$
and for any $k\leq i$ we write $\tau(\bar{\omega}_{i})\leq k\text{ if }(\bar{\omega}_{i}\omega_{i+1}\dots\omega_{N})\in\{\tau\leq k\}$
for all possible paths $(\bar{\omega}_{i}\omega_{i+1}\dots\omega_{N})$,
where $\omega_{j}\in\{u,d\}$ for all $i+1\leq j\leq N$. 
\end{Definition}
It is convenient to define the respective non-decreasing sequences
of stopping times $\{\tau_{i}\}_{i=0,1,\dots,N}$ and $\{\bar{\tau}_{i}\}_{i=0,1,\dots,N}$
with $\tau_{i}:=\tau\vee i$ and $\bar{\tau}_{i}:=\tau_{i}\wedge N$
for $i\in\mc{I}_{N}$. For $0\leq j\leq i\leq N$ and $k\in\{i,i+1,\dots,N\}\cup\{\infty\}$,
by the Markov property of $\{W_{i}\}$ we have 

\begin{equation}
\Q(\tau_{j}=k|\mc{F}_{i})=H(i,j,k,W_{i^{-}}),\label{eq:Markov prop Qtau-1}\end{equation}

\noindent
where

\[
H(k\wedge N,j,k,x)=\begin{cases}
\mathbf{1}_{\{x>0,w(x)=0\}} & k\leq N,\\
\mathbf{1}_{\{w(x)>0\}} & k=\infty,\end{cases}\] 

\noindent
and for $i\vee1\leq l<k\wedge N$

\[
H(l,j,k,x)=\begin{cases}
pH(l+1,j,k,w(x)u)+qH(l+1,j,k,w(x)d) & x>0,\\
0 & x=0.\end{cases}\] 

\noindent
For $i=0$, we have

\[
H(0,0,k,x)=pH(1,0,k,xu)+qH(1,0,k,xd).\] 

\noindent
If $\tau=\infty$ the contract matures with a positive account value
at time $N\delta t=T$ and the option is not exercised, that is the guarantee expires worthless.

Since the value processes at each time point are ex-fees and ex-withdrawals,
the component $(G-W_{\tau^{-}}e^{-\bara})\geq0$ is the rider payment
made immediately at trigger time. For any period $i$, the net rider
payout at time $i\delta t$ is

\begin{equation}
\label{eq:11b}
(G-W_{i^{-}}e^{-\bara})^{+}-W_{i^{-}}(1-e^{-\bara}).
\end{equation}

\noindent
Therefore, by the risk neutral pricing formula the value at time $i$ of the rider value process is given by

\begin{align}
U_{i} & =E_{\Q}\left[\sum_{j=i+1}^{N}e^{-\barr(j-i)}\left[\left(G-W_{j^{-}}e^{-\bara}\right)^{+}-W_{j^{-}}\left(1-e^{-\bara}\right)\right]|\mc{F}_{i}\right]\nonumber \\
 & =E_{\Q}\Bigr[\left(G-W_{\bar{\tau}_{i}^{-}}e^{-\bara}\right)e^{-\barr(\bar{\tau}_{i}-i)}\mathbf{1}_{\{i+1\leq\bar{\tau}_{i}\}}+\sum_{m=\bar{\tau}_{i}+1}^{N}Ge^{-\barr(m-i)}%
  -\sum_{m=i+1}^{\bar{\tau}_{i}}e^{-\barr(m-i)}W_{m^{-}}\left(1-e^{-\bara}\right)|\mc{F}_{i}\Bigl]\label{eq:Udiscrte} \end{align}

\noindent
for $i\in\mc{I}_{N-1}$. The terminal value is $U_{N}=0$.  Note that equation~(\ref{eq:Udiscrte}) is the discrete-time binomial model analogue of equation~(10) in \citet{Hyndman-Wenger-Decomp}.

By the Markov property for $\{W_{i}\}$ we have $U_{i}=u(i,W_{i})$,
where $u:\mc{I}_{N}\times\mb{R}_{+}\mapsto\mb{R}$ is
defined by%
\footnote{There is some abuse of notation with $u$ referring both to the up-movement
in the binomial model and to the rider value function. However, it
is always clear from the context whether we are referring to the constant
value or to the rider value function.%
}

\begin{equation}
u(i,x)=\begin{cases}
0 & i=N,\\
e^{-\barr}[pu^{-}(i+1,xu)+qu^{-}(i+1,xd)] & 0\leq i<N,\end{cases}\label{eq:UiShortVersion-1}\end{equation}

\noindent
where  $u^{-}:\mc{I}_{N}^{+}\times\mb{R}_{+}\mapsto\mb{R}$
is defined by 

\begin{equation}
u^{-}(i,x)=u(i,w(x))+(G-xe^{-\bara})^{+}-x(1-e^{-\bara}),\label{eq:u-}\end{equation}

\noindent
 and $w(x)$ is provided by \prettyref{eq:w(x)}. The function $u^{-}(i,x)$
represents the rider value at time point $i$ cum-fees and cum-withdrawals,
where $x$ is the AV before fees and withdrawals are deducted.

Since the policyholder and insurer valuation equations \prettyref{eq:V_0Discrete} and \prettyref{eq:Udiscrte}
are the respective discrete-time versions of equations (7) and (10) of \citet{Hyndman-Wenger-Decomp}
we expect that the relationship between the policyholder and insurer valuation perspectives carries over 
from the continuous time case.  That is, we expect in the binomial model that the value of the complete contract can also be decomposed
as the sum of the value of the account value and the value of the guarantee.
Indeed, this can be shown directly from \prettyref{eq:V_0Discrete} and \prettyref{eq:Udiscrte}. 
We provide an alternative proof applying backward induction to the functions $v(i,x)$ and $u(i,x)$.
\begin{Theorem}
\label{thm:V=00003DU_WBin}
Under \prettyref{ass:bin no lapse}, for
all $\alpha\geq0$ we have

\[
V_{i}=U_{i}+W_{i}\] 

\noindent
 for all $i=0,1,\dots,N$. 
\end{Theorem}
\begin{proof}
See Appendix~\ref{Proofs}.
\end{proof}
One advantage of the discrete-time binomial framework is that it allows us to give explicit hedging strategies for replicating contingent claims.  We next discuss hedging the GMWB rider.

\subsection{\label{sub:Hedging}Hedging}

Consider the no-hedging strategy where fee revenues are
invested at rate $r$ in a money market account  and at time $\tau$, if $\tau<\infty$,
the rider payoff is paid from this account. The $\mc{F}_{\bartauzero}$-measurable
random variable $\mc{C}_{\bartauzero}$ measures the total
cost of the rider to the insurer over the contract lifespan, discounted
to time zero, when hedging is not used. Denote the periodic fees received at time-step $i$ by $F{}_{i}$.
Then we have, by the definition of the contract, that 
 $F{}_{i}:=W_{i^{-}}(1-e^{-\bara})$
for $i\in\mc{I}_{N}^{+}$ and $F_{0}=0$. We have

\[
\mc{C}{}_{\bartauzero}=e^{-\barr\bartauzero}\left[\left(G-(W_{\bartauzero^{-}})e^{-\bara}\right)^{+}+Ga_{\lcroof{N-\bartauzero}}-\sum_{i=1}^{\bartauzero}F{}_{i}\times e{}^{\barr(\bartauzero-i)}\right].\] 

\noindent
Note that $U_{0}=E_{\Q}[\mc{C}_{\bartauzero}]$, but
we are concerned with the pathwise results of $\mc{C}_{\bartauzero}$
in relation to the outcomes resulting from a dynamic hedging strategy. 

The insurer establishes a hedging portfolio,
which attempts to replicate the rider so that any rider claims can
be fully paid out by the portfolio. The party managing the rider risk
does not have access to the account value funds to mitigate any risk,
rather the only sources of revenue are the rider fees. Denoting the
replicating portfolio by $\{X_{i}\}$, the objective is to have $X_{i}=U_{i}$
for all $i$ in a pathwise manner.

Define the adapted portfolio process
$\{\Delta_{i}\}_{0\leq i\leq N-1}$. On each time interval $[i\delta t,(i+1)\delta t)$
until maturity and for all outcomes the replicating portfolio 
maintains a position of $\Delta_{i}(\omega_{1}\dots\omega_{i})$ units
in $S$. Using the Markov property of $\{W_{i}\}$ we define $\Delta_{i}:=\Delta(i,W_{i},S_{i})$,
 where $\Delta:\mc{I}_{N-1}\times\mb{R}_{+}\times\mb{R}_{+}\mapsto\mb{R}$
is given by

\begin{equation}
\Delta(i,x,y)=\frac{u^{-}(i+1,ux)-u^{-}(i+1,dx)}{uy-dy}.\label{eq:DeltaBinNolapse}\end{equation}

\noindent
This indicates that $\Delta_{i}=0$ for $\tau\leq i\leq N-1$ as no uncertainty
remains. By the nature of the rider as an embedded put-like option,
$\Delta$ will always take non-positive values corresponding to short
positions in $S$.  Any positive (negative) portfolio cash balance
is invested in (borrowed from) the money market at rate $r$.

Beginning with initial capital $X_{0}=x_{0}\in\mb{R}$, the replicating
portfolio $\{X_{i}\}$ follows

\begin{equation}
X_{i}=\left(X_{i-1}-\Delta_{i-1}S_{i-1}\right)e^{\barr}+\Delta_{i-1}S_{i}+F_{i}-(G-W_{i^{-}}e^{-\bara})^{+}\label{eq:Xi}\end{equation}

\noindent
for $i\in\mc{I}_{N}^{+}$. Over any period the change in the
portfolio value of $(X_{i}-X_{i-1})$ consists of the sum of four components:
a) the return in the money market earned on both the prior portfolio
balance and the proceeds from the shorted stock $(X_{i-1}-\Delta_{i-1}S_{i-1})(\exp{(\barr)}-1)$;
b) the capital gain or loss on the shorted stock $(S_{i}-S_{i-1})\Delta_{i-1}$;
c) the end of period rider fees $F_{i}$; and d) the negative of that
period's rider claim (if any), paid at the end of the period and given
by $(G-W_{i^{-}}\exp{(-\bara)})^{+}$. Note that if the static
hedging strategy $\Delta\equiv0$ is used then $X_{N}\exp{(-\barr N)}=-\mc{C}_{\bartauzero}$.
That is, we  obtain the no-hedging result.

Similar to \citet[Theorem 2.4.8]{RefWorks:33} the portfolio process given by (\ref{eq:DeltaBinNolapse})
replicates the rider value.  The proof is omitted as we shall prove a more general result after we generalize the model to include lapses.
\begin{Theorem}
\label{thm: Hedg GMWB}Under \prettyref{ass:bin no lapse}, if the
fee $\alpha$ is charged and the initial capital is $x_{0}=U_{0}(P,\alpha,g)$,
then an insurer who maintains the replicating portfolio $X_{i}$ by
following the portfolio process prescribed by \prettyref{eq:DeltaBinNolapse}
will be fully hedged. That is,

\[
X_{i}=U_{i}\] 

\noindent
 for $i\in\mc{I}_{N}$.
\end{Theorem}
\begin{Remark}
In particular, if $\tau\leq N$ then $X_{\tau}=G\times a_{\lcroof{N-\tau}}$.
When $\faira$ is charged we have $U_{0}=0$ and no initial capital
is required for the replicating portfolio. The rider is different
from standard financial options in that there is no upfront cost
to finance the hedge but rather it is self-financed through periodic
contingent fees. If the fee charged is not the fair fee ($\alpha\neq\faira)$,
then the insurer must make an initial deposit to the hedging portfolio
if $\alpha<\faira$ or may consume from the portfolio at time
zero if $\alpha>\faira$. The insurer can justify a lower
fee by either depositing capital into the portfolio and selling the
policy at a loss or by charging an initial fee per unit premium at
time zero to the insured.
\end{Remark}

\section{\label{sec:Binlapses}Optimal Stopping and Surrenders}

We next extend the binomial pricing model to include the possibility of
early surrenders by modifying \prettyref{ass:bin no lapse} to include the 
following assumption. 
\begin{Assumption}
\label{ass:Bin Lapses} Under the static withdrawal strategy the policyholder
receives $G=gP\delta t$ each time period. We set $T:=1/g$. 
At the end of each period the pro-rated rider fee is first deducted
and then the periodic withdrawal is subtracted. 
We restrict $r>0$ and denote $\barr:=r\delta t$ and $\bar{\alpha}:=\alpha\delta t$.
Surrenders occur at the end of any
time period, after the fees and withdrawals have been deducted. For
valuation purposes, the end of period time point is considered ex-post
fees and withdrawals but ex-ante surrenders. 
\end{Assumption}
Let $k^{a}:\{0,1,\dots T\}\mapsto[0,1]$ be the non-increasing function
describing the surrender charge schedule, satisfying $k_{0}^{a}>0$
and $k_{T}^{a}=0$. The surrender charge rate $k_{i}^{a}$ is applied
for surrenders during time $[i,i+1)$. We denote the corresponding
function for the surrender charge rate upon surrender at the end of
period $i$ by $k:\{0,1,\dots,N\}\rightarrow[0,1]$. Then $k_{i}=k_{\lfloor i\delta t\rfloor}^{a}$.
Similar to the continuous-time model of \citet{Hyndman-Wenger-Decomp} for 
all $i\in\mc{I}_{N}$ we have 

\begin{align}
V_{i} & =\max_{\eta\in\mb{L}_{i}}V_{i}^{\eta}=\max_{\eta\in\mb{L}_{i,\bar{\tau}_{i}}}V_{i}^{\eta},\label{eq:v0max}\end{align}

\noindent
where 

\begin{equation}
V_{i}^{\eta}=E_{Q}\left[Ga_{\lcroof{\eta-i}}+W_{\eta}(1-k_{\eta})e^{-\barr(\eta-i)}|\mc{F}_{i}\right],\label{eq:Vilapse}\end{equation}

\noindent
$\mb{L}_{i}$ is the set of $\mb{F}-$adapted
stopping times taking values in $\{i,i+1,\dots,N\}$, and $\mb{L}_{i,\bar{\tau}_{i}}$ is the set of $\mb{F}-$adapted
stopping times taking values in $\{i,i+1,\dots,N\}$ subject to the
constraint $\eta<\bar{\tau}_{i}$ or $\eta=N$. Recall that $\bar{\tau}_{i}$
is the trigger time assuming no lapses.

With the objective of classifying the optimal surrender policy we
introduce some notation. For any $0\leq i\leq N,$ define a rescaled
filtration $\mb{F}^{i}=\{\mc{F}_{j}^{i}:=\mc{F}_{j+i};0\leq j\leq N-i\}$.
For any $\eta\in\mb{L}_{i}$ define 

\begin{equation}
Y^{\eta,i}:=\left\{ Y_{j}^{\eta,i}=e^{-\barr((j+i)\wedge\eta)}V_{(j+i)\wedge\eta}^{\eta}+Ga_{\lcroof{(j+i)\wedge\eta}}\right\} _{0\leq j\leq N-i},\label{eq:mtgale}\end{equation}

\noindent
then $Y^{\eta,i}$ is a $(\Q,\mb{F}^{i})$ martingale.
Define the surrender policy $\tilde{\eta}$ by

\begin{equation} 
\tilde{\eta}_{i}:=\min\{j\geq i;\ V_{j}=W_{j}(1-k_{j})\}\leq N . \label{eq:etahat}\end{equation}

\noindent
A policyholder following the surrender strategy given by equation~(\ref{eq:etahat}) lapses 
at the first time  valuation of the contract, from the policyholder's perspective, is equal to 
the account value less the surrender charge.
This is similar to the classical result from American contingent claims theory which gives 
that $\tilde{\eta}_{i}$ is optimal in the sense that $V_{i}=V_{i}^{\tilde{\eta}_{i}}$
(proving this in our context is straightforward based on \citet[p.35]{RefWorks:63}
but requires \prettyref{eq:mtgale}). That is, $\tilde{\eta}_{i}$
is an optimal surrender policy for the insured to follow going forth
from time $i\delta t$, given the current market state and no prior
surrender.  %

The backward induction (risk-neutral pricing) algorithm is constructed to evaluate $V$ on a
binomial tree. By the Markov property for $\{W_{i}\}$ we have $V_{i}=v(i,W_{i})$,
where $v:\mc{I}_{N}\times\mb{R}_{+}\mapsto\mb{R}_{+}$
is given recursively as

\[
\begin{cases}
v(N,x)=x(1-k_{N})=x,\\
v(i,x)=\max\{(G+pv(i+1,w(ux))+qv(i+1,w(dx)))e^{-\barr},x(1-k_{i})\}.\end{cases}\] 

\noindent
When solving for $\faira$ we may write 

$$v(0,P)=[G+pv(1,w(uP)+qv(1,w(dP))]e^{-\barr}$$

\noindent
since $k_{0}>0$. 

Consider the rider value $U$ by extending equation~(\ref{eq:Udiscrte}) to incorporate the option to surrender and 
receive the payoff $k_{\eta}W_{\eta}$ at surrender.  Then, at time $i$ the rider value is given by

\begin{equation}
U_{i}:=\max_{\eta\in\mb{L}_{i,\bar{\tau}_{i}}}U_{i}^{\eta} \label{eq:ULapseBin}\end{equation}

\noindent
where 

$$
U_{i}^{\eta} = E_{\Q}[\negmedspace\sum_{j=i+1}^{\eta}\negmedspace e^{-\barr(j-i)}[(G-W_{j^{-}}e^{-\bara})^{+}-W_{j^{-}}(1-e^{-\bara})]-e^{-\barr(\eta-i)}k_{\eta}W_{\eta}|\mc{F}_{i}]
$$

\noindent
using the convention that $\sum_{j=i+1}^{i}(\cdot)=0$. Note that equation~(\ref{eq:ULapseBin}) is the discrete-time analogue of the rider value in the continuous-time model given in equation~(14) of \citet{Hyndman-Wenger-Decomp}.

The value of the option to surrender, $L$, is the difference between the rider value when lapses are allowed and the rider value without lapses. That is, define $L_{i}:=U_{i}-U_{i}^{NL}\geq 0$,
where $U_{i}^{NL}$ is the rider value in the no-lapse case \prettyref{eq:Udiscrte}.
Then at time $i$ the value of the option to lapse is given by

\begin{equation}
L_{i}=\max_{\eta\in\mb{L}_{i,\bar{\tau}_{i}}}L_{i}^{\eta},\label{eq:BinLapse}\end{equation}

\noindent
where 

\begin{equation*}
L_{i}^{\eta}=E_{\Q}[\sum_{j=\eta+1}^{N}e^{-\barr(j-i)}[W_{j^{-}}\left(1-e^{-\bara}\right)-\left(G-W_{j^{-}}e^{-\bara}\right)^{+}]-e^{-\barr(\eta-i)}k_{\eta}W_{\eta}|\mc{F}_{i}].
\end{equation*}

\noindent
Note that equation~(\ref{eq:BinLapse}) is the discrete time analogue of the value of the option to lapse in the
continuous-time model given by equation~(15) of \citet{Hyndman-Wenger-Decomp}.

Write $U_{i}=u(i,W_{i})$,
 where $u: \mc{I}_{N}\times\mb{R}_{+}\mapsto\mb{R}$
is recursively defined by

\[
\begin{cases}
u(N,x)=-k_{N}x=0,\\
u(i,x)=\max\{e^{-\barr}[pu^{-}(i+1,ux)+qu^{-}(i+1,dx)],-k_{i}x\},\end{cases}\] 

\noindent
 and $u^{-}:\mc{I}_{N}^{+}\times\mb{R}_{+}\mapsto\mb{R}$
follows

\begin{equation}
u^{-}(i,x)=u(i,w(x))+(G-xe^{-\bara})^{+}-x(1-e^{-\bara}).\label{eq:uminuslapse}\end{equation}

\noindent
Denoting the rider value function in the no-lapse model from \prettyref{eq:UiShortVersion-1}
by $u^{NL}(i,x)$, we have $L_{i}=l(i,W_{i})$, where $l:\mc{I}_{N}\times\mb{R}_{+}\mapsto\mb{R}_{+}$
is given by

\[
\begin{cases}
l(N,x)=-k_{N}x=0,\\
l(i,x)=\max\{e^{-\barr}(pl(i+1,w(ux))+ql(i+1,w(dx))),-u^{NL}(i,x)-k_{i}x\}.\end{cases}\] 

\noindent
Note that $u^{NL}(i,0)\geq0$ which implies the
boundary condition $l(i,0)=0$. Once the rider is triggered, early
surrender is suboptimal since any remaining
guarantee is forfeited upon surrender. 

In the case of lapses we may extend \prettyref{thm:V=00003DU_WBin} to 
decompose the value of the complete contract into the 
sum of the account value and the value of the guarantee. 
\begin{Theorem}
\label{thm:VUWLBin}Under \prettyref{ass:Bin Lapses}, for all $\alpha\geq0$
and for all $i\in\mc{I}_{N}$, we have 

\begin{equation}
V_{i}=U_{i}+W_{i},\label{eq:VUW(Lapse)}\end{equation}

\noindent
or equivalently 

\begin{equation}
V_{i}=L_{i}+U_{i}^{NL}+W_{i}.\label{eq:v+l+u+wbinlap}\end{equation}

\noindent
\end{Theorem}
\begin{proof}
Equation \prettyref{eq:VUW(Lapse)} can be proved using backward induction
on the recursive functions $v$ and $u$, similar to \prettyref{thm:V=00003DU_WBin}.
We omit the details.\end{proof}
Note that \prettyref{thm:VUWLBin} is the discrete-time analogue of the continuous time decomposition
given in \citet[Theorem 7]{Hyndman-Wenger-Decomp}.

As in the no-lapse case an advantage of the discrete-time binomial model is that we are easily able to hedge the guarantee.

\subsection{Hedging with lapses}
We next extend the standard hedging results for American derivatives (see \citet[Theorem 4.4.4]{RefWorks:33})
by incorporating the complication of the periodic revenues and rider
claims We show that the insurer can perfectly hedge the rider risk
by maintaining the appropriate replicating portfolio.  The adapted portfolio process $(\Delta_{i})_{0\leq i<N}$ remains unchanged from  \prettyref{eq:DeltaBinNolapse}, except that $u^{-}(i,x)$ is given by \prettyref{eq:uminuslapse}. Furthermore, the insurer may have positive consumption under suboptimal surrender behaviour.  

Define the consumption process $C=\{C_{i}\}_{0\leq i<N}$ by
$C_{i}:=c(i,W_{i})$  where
$c:\mc{I}_{N-1}\times\mb{R}_{+}\mapsto\mb{R}_{+}$ is
given by

\begin{equation}
c(i,x):=v(i,x)-[pv(i+1,w(ux))+qv(i+1,w(dx))+G]e^{-\barr}\geq0.\label{eq:consumption-1}\end{equation}

\noindent
The consumption process $\{C_{i}\}$ represents the additional cash flow received each time
a policyholder behaves sub-optimally by not surrendering. We can explicitly classify suboptimal behaviour by defining a sequence of stopping times.

With $\tilde{\eta}_{i}$ defined as in equation~\prettyref{eq:etahat} let $\tilde{\eta}^{0}:=\tilde{\eta}_{0}$
and for $1\leq j\leq m$ we denote

\[
\tilde{\eta}^{j}=\tilde{\eta}_{z_{j}},\] 

\noindent
where $z_{0}=0$, $z_{j}=(\tilde{\eta}^{j-1}+1)\wedge N$, and $m=\min\{i;\tilde{\eta}^{i}=N\text{ a.s.}\}$.
Then  we may characterize, in terms of $\{\tilde{\eta}^{j}\}$,
precisely when $C$ will be strictly positive. We have $C_{\tilde{\eta}^{j}}>0$
for all $0\leq j<M:=\min\{b;z_{b}=N\}\leq m$, where $M$ is a random
variable. Otherwise $C_{i}=0$.

There is a fine distinction between $C_{\tilde{\eta}^{j}}$ and $L_{\tilde{\eta}^{j}}$
for all $j<M$. Consider the two surrender strategies of $\tilde{\eta}^{j+1}$
and $\eta=N$. The first strategy corresponds to surrendering at the
next best time after $\tilde{\eta}^{j}$ and the latter strategy is
equivalent to never surrendering early. Then $C_{\tilde{\eta}^{j}}=V_{\tilde{\eta}^{j}}-V_{\tilde{\eta}^{j}}^{\tilde{\eta}^{j+1}}$
but $L_{\tilde{\eta}^{j}}=V_{\tilde{\eta}^{j}}-V_{\tilde{\eta}^{j}}^{NL}$.
At any time when it is optimal to surrender immediately, $C$ provides
the marginal value from surrendering now instead of at the next optimal
time, whereas $L$ is the marginal value from acting now instead of
at maturity.

By \prettyref{thm:V=00003DU_WBin} and \prettyref{thm:VUWLBin} it
follows that $V_{\tilde{\eta}^{j}}^{\tilde{\eta}^{j+1}}=U_{\tilde{\eta}^{j}}^{\tilde{\eta}^{j+1}}+W_{\tilde{\eta}^{j}}$
and $V_{\tilde{\eta}^{j}}=U_{\tilde{\eta}^{j}}+W_{\tilde{\eta}^{j}}$.
Therefore $C$ can be written in terms of $U$ as

\begin{equation}
c(i,x)=u(i,x)-[pu^{-}(i+1,ux)+qu^{-}(i+1,dx)]e^{-\barr}.\label{eq:Consumption}\end{equation}

\noindent

Beginning with $X_{0}=x_{0}$, the replicating
portfolio is constructed  recursively forward in time taking into consideration fee revenues, consumption, and rider claim payments. For all $i\in\mc{I}_{N}^{+}$
we have

\begin{equation}
X_{i}=\left[X_{i-1}-\Delta_{i-1}S_{i-1}-C_{i-1}\right]e^{\barr}+\Delta_{i-1}S_{i}+F_{i}-(G-W_{i^{-}}e^{-\bara})^{+}.\label{eq:Xilapse}\end{equation}

\noindent

\begin{Theorem}
\label{thm: Hedg GMWB-1}Under \prettyref{ass:Bin Lapses}, if the initial capital is $x_{0}=U_{0}$, then an insurer who maintains the replicating portfolio $X_{i}$ defined
by \prettyref{eq:Xilapse} and liquidates the portfolio
either upon early surrender (if any) or at time point $N$ will be
fully hedged throughout the contract lifespan. That is, for
all $i\in\mc{I}_{N}$ and all surrender strategies

\[
X_{i}=U_{i}.
\] 

\noindent
\end{Theorem}
\begin{proof}
See Appendix~\ref{Proofs}.
\end{proof}
\begin{Remark}
Assuming the insured follows
the optimal surrender strategy $\tilde{\eta}_{0}$, then $X_{\tilde{\eta}_{0}}=U_{\tilde{\eta}_{0}}$
and on $\{\tilde{\eta}_{0}<\bartauzero\}$ we have that $X_{\tilde{\eta}_{0}}=U_{\tilde{\eta}_{0}}=-k_{\tilde{\eta}_{0}}W_{\tilde{\eta}_{0}}$,
whereas $X_{N}=U_{N}=0$ on $\{\tilde{\eta}_{0}=N\}$. Under this strategy there is no
consumption. However, if the insured allows the first optimal surrender time
$\{\tilde{\eta}_{0}<\bartauzero\}$ to elapse, then the insurer
will consume $C_{\tilde{\eta}_{0}}$ and the remaining portfolio is
still sufficient to hedge the contract over the remaining lifespan.
If the insured allows the next optimal surrender time $\{\tilde{\eta}_{0}<\tilde{\eta}^{1}<\bartauzero\}$
to elapse, if it exists, then the insurer consumes an additional $C_{\tilde{\eta}^{1}}$
and this continues until the earlier of trigger or time point $N$.

Finally suppose the insured surrenders at a suboptimal time. For a
given path $\bar{\omega}_{N}$, surrender occurs at a time point $i\neq\tilde{\eta}^{j}$
for all $0\leq j\leq M(\bar{\omega}_{N})$. Then the insured receives
$W_{i}(1-k_{i})$ and in turn foregoes $V_{i}-W_{i}(1-k_{i})>0$ of
value. The insurer's portfolio value is $X_{i}+k_{i}W_{i}>0$ and
the insurer has a positive consumption. Indeed by \prettyref{eq:VUW(Lapse)}
we have $V_{i}-W_{i}(1-k_{i})=U_{i}+W_{i}k_{i}>0$, but $X_{i}=U_{i}$. 
\end{Remark}

With the explicit recursive formulae for pricing and hedging the contract we may consider the
implementation of the binomial model and its performance relative to
the theoretical results presented and other modeling approaches which have appeared in the literature.  
We first briefly address computational considerations of the
binomial model.

\section{\label{sec:Comput}Computational Considerations}

Computational applications of the binomial model for the GMWB
rider face two specific challenges. The binomial tree for the account
value process is non-recombining and the riders have significantly
longer durations in contrast to the usual European and American equity
options which typically have durations not exceeding one year. The
withdrawal rate $g$ can be expected to range from 5\% to 10\% corresponding
to maturities of 10 to 20 years. 
If the value processes in the binomial world is
to provide an accurate approximation of the value processes in the
continuous-time model of \citet{Hyndman-Wenger-Decomp}
$\delta t$ must be significantly smaller than one.

The backward induction (tree) algorithm (referred to as Method A) for calculating $V_{0}$ involves arrays of size $2^{N}$ to record $V_{N}$ for all nodes in the final period. In contrast, for recombining trees the array
size needed is only $N+1$. For $g=5\%$ the binomial tree will contain $2^{20}>10^{6}$ nodes
in the final period with just one time-step per year. Method A requires too much memory for
small values of $\delta t$. 

We will show that in the no-lapse model we can directly calculate $v(i,x)$ without using trees and avoid the strain on memory capacity from storing the large arrays of data. This direct approach (Method B) uses an algorithm which loops through each path requiring minimal memory. We will see shortly that despite being able to eliminate a subset of the paths from the looping process this method is significantly slower than Method A. Although Method B
enables using marginally smaller $\delta t$ values, we quickly run
into time constraints as the number of paths grows at $O(2^{N})$.

We will then introduce an approximation method which uses the backward induction (tree) approach while easing the memory strain. This retains the flexibility to model the GMWB both with and without lapses. Further it avoids the time constraints with Method B. 

The terminal AV can be expressed directly as:

\begin{align}
W_{N} & =\max\left[\xi_{N}e^{-\bara}\left(\xi_{N-1}e^{-\bara}\left(...\left(\xi_{2}e^{-\bara}\left(P\xi_{1}e^{-\bara}-G\right)-G\right)...\right)-G\right)-G,0\right]\nonumber \\
 & =\max\left[0,Pe^{-\bara N}\prod_{i=1}^{N}\xi_{i}-G\sum_{i=0}^{N-1}e^{-\bara i}\prod_{j=N-(i-1)}^{N}\xi_{j}\right],\label{eq:W_Ndirect}\end{align}

\noindent
where the convention $\prod_{N+1}^{N}(\cdot)=1$ is used.
Applying the reversal technique from \citet{RefWorks:43}, which is justified by the exchangeability
property of the sequence $\{\xi_{i}\}_{i=1}^{N}$, and considering the
reversed sequence which is equal in distribution, it follows that

\[W_{N}^{x,M}\overset{d}{=}\max\left[0,xZ_{N-M}-G\sum_{i=0}^{N-M-1}Z_{i}\right],\label{eq:asianZ}\] 

\noindent
where $M<N$ and $\{Z_{i}\}$ is the account value process when there are no withdrawals,
beginning with $Z_{0}=1$. In particular, with $M=0,$ $x=P$, and $G=P/N$ we obtain that  $V_{0}$ can be expressed
as a floating-strike Asian call option on $\{Z_{i}\}$ plus a term
certain component, as pointed out by \citet{RefWorks:43}. 

Many of the terminal nodes in the tree for $\{W_{i}\}$ will be zero
as a result of the periodic withdrawals, fees, and possible negative
returns on $S$. Consider the recombining tree for $\{Z_{i}\}$ with
$N+1$ nodes for period $N$. At each node, for each path leading
to it the average must be computed to calculate $W_{N}$. Suppose
that for some $i\leq N$ we have $W_{N}=0$ on all paths with $i$
jumps of $u$ and $N-i$ jumps of $d$. Then $W_{N}=0$ for all paths
with less than $i$ jumps of $u$. Consequently, once we reach a node
on the tree for $Z$ such that $W_{N}=0$ for all paths, no further
paths need be considered.

There is an efficient permutation function in \CC, \textit{next\_permutation},
which quickly loops through all distinct paths having $i$ jumps of
$u$ and $N-i$ jumps of $d$. By looping through each node and its respective paths we can avoid the exponential growth in memory
storage, although we show in our numerical results that the run-time
will increase significantly. By \prettyref{eq:V_0Discrete}, with
$\zeta:=N-m$ we can write

\begin{align}
v(m,x) &= G a_{\lcroof{\zeta}}+e^{-\barr\zeta}\sum_{k=0}^{A_{0}}p^{\zeta-k}q^{k}\sum_{\Xi_{\zeta,k}}\Bigg(xe^{-\bara \zeta}u^{\zeta-k}d^{k} %
       - G \sum_{i=0}^{\zeta-1}e^{-\bara i}\prod_{j=1}^{i}\omega_{j}\Bigg)^{+}, \label{eq:Vonovector}
\end{align}

\noindent 
where $\Xi_{\zeta,k}$ is the set of ${\zeta \choose k}$ unique
permutations of a path with $\zeta-k$ up and $k$ down movements
and $A_{0}$ is the first value of $k$ for which the summand produces
zero.

\citet{RefWorks:83} developed an approximation method to value path-dependent
financial options on a binomial lattice in a more efficient manner.
The key idea is to use only a representative set of averages at each
node and apply linear interpolation in the backwards induction scheme.
\citet{RefWorks:84} discuss several drawbacks of the \citet{RefWorks:83} method
and  propose a different approximation method and in particular provide the details for pricing fixed-strike European and American Asian
call options. Numerical results show
convergence for European Asian calls while American Asian calls do
not perform as well, converging at a much slower rate. The method is easily modified for any option
payoff which depends on a valid function of the asset price path.

The options considered by \citet{RefWorks:84} have significantly
shorter maturities compared to the GMWB riders. The method reduces
the number of contract values considered in the backwards induction
scheme from $O(2^{N})$ to $O(N^{4})$. In our work, memory constraints
limited the number of time steps in the binomial trees to $N=28$
but with this method we can consider up to $N=128$ time-steps. We
briefly describe the approximation method applied to GMWBs with lapses
but refer the reader to \citet{RefWorks:84} for more details on the scheme. 

Using equation~\prettyref{eq:W_Ndirect} we can rewrite the value of the contract to the 
policyholder given by equation~\prettyref{eq:v0max} as

\begin{equation}
V_{0}=\max_{\eta\in\mb{L}_{0}}E_{Q}\left[Ga_{\lcroof{\eta}}+P\max\left(Z_{\eta}\left(1-\frac{1}{N}\sum_{i=1}^{\eta}\frac{1}{Z_{i}}\right),0\right)\left(1-k_{\eta}\right)e^{-\barr\eta}\right],\label{eq:V0LapseAsain-1}\end{equation}

\noindent
where

\[
Z_{n}=\prod_{i=1}^{n}e^{-\bara}\xi_{i}=e^{-\bara n}\frac{S_{n}}{S_{0}}.\] 

\noindent
Therefore, 

$$
  V_{i}=v(i,Z_{i},\sum_{j=1}^{i}Z_{j}^{-1}),
$$ 

\noindent
where $v:\mc{I}_{N}\times\mb{R}_{+}\times\mb{R}_{+}\mapsto\mb{R}_{+}$
is defined recursively backward in time by

\[
v(N,x,y)=P\max\left(x\left(1-\frac{1}{N}y\right),0\right)\] 

\noindent
for $i=N$ and 

\begin{align*}
v(i,x,y) &= \max\bigg[\Bigl[G+pv\negmedspace\left(i+1,xue^{-\bara},y+\left(xue^{-\bara}\right)^{-1}\right)\\
 & +qv\negmedspace\left(i+1,xde^{-\bara},y+\left(xde^{-\bara}\right)^{-1}\right)\Big]e^{-\barr},x\left(1-\frac{1}{N}y\right)\left(1-k_{i}\right)\bigg]
\end{align*}

\noindent
for $0\leq i<N$.

Let $(i,j)$ denote the node reached by $j$ up-movements and $(i-j)$
down-movements in the recombining tree for $Z$. We write $z(i,j)$
for the value of $Z$ at node $(i,j)$. For each node, we construct
a set of $j(i-j)+1$ representative averages, where the terminology of \textit{average} is used even though we do not
divide by $i+1$. This set is a subset of the complete set of ${i \choose j}$ averages
for the paths at that node. Denote the first (and lowest) element
by $A(i,j,1)$ where

\[
A(i,j,1)=\sum_{h=0}^{j}\left(ue^{-\bara}\right)^{-h}+\left(ue^{-\bara}\right)^{-j}\sum_{h=1}^{i-j}\left(de^{-\bara}\right)^{-h}.\] 

\noindent
This average is taken along the path beginning with $j$ up-movements
of $u$ and followed by $(i-j)$ down-movements of $d$. Excluding
the initial point and terminal point we find the highest point of
$\{S_{i}\}$ along the path (if there are more than one such points,
select the first one) and substitute that node with the node directly
below it in the $\{Z_{i}\}$ tree to obtain a new path and take its
average. This is repeated $j(i-j)$ times to obtain the set $A(i,j)=\{A(i,j,k);1\leq k\leq j(i-j)+1\}$.
The final path considered will be the one with $(i-j)$ down-movements
followed by $j$ up-movements. None of the previous paths are allowed
to be below this path.

When working with the function $v$ on the tree for $Z$ and applying
backward induction, linear interpolation is used whenever the
computed average is not in the representative set for that node. This is done by considering the two nearest elements of the set,
one on each side of the computed average (see \citet{RefWorks:83} and \citet{RefWorks:84} for details). The scheme
from \citet{RefWorks:84} has the benefit that linear interpolation
is not needed for many of the computations of $v$.

For the framework in \citet{RefWorks:84}, whether the algorithm begins
with the path giving the highest average, selects paths in the described
manner, and stops when the path giving the lowest average is obtained,
or vice versa, the same set of averages are obtained. This symmetry
is a result of the underlying asset changing by factors of $u$ and
$d$, where $ud=1$. However, this symmetry does not hold in our model
because the process $Z$ changes by factors of $ue^{-\alpha}$ and
$de^{-\alpha}$. For example, an up-move followed by a down-move does
not return $Z$ to its initial value. The downward trend of the $Z$-tree
complicates the approximation algorithm. Consequently, the sets $A(i,j)$
will change depending on whether the lowest or highest path is initially
considered.

\section{\label{sec:Num-Res-No-Mort}Numerical Results: Excluding Mortality Risk}

Beginning with the no-lapse case, we provide numerical results comparing
our model to previous results in the literature, which excluded mortality risk, and find that even
with large values for $\delta t$ our simple model is a reasonable
approximation of more complex models. Moreover, the discrete-time binomial model
allows us to analyze the hedging results and the effect
of the parameters on the losses when hedging is not implemented. 

\subsection{\label{sub:Numerical-results-alpha}The Fair Rider Fee}

The bisection algorithm is used to numerically solve for $\faira$ given by Definition~\ref{defn:fair-alpha}.
Define $f:\mathbb{R}_{+}\mapsto\mathbb{R}_{+}$
by $f(\alpha)=V_{0}(P,\alpha,g)-P$. Then $f(\alpha^{\star})=0$ by Definition~\ref{defn:fair-alpha}. 
We use $P=100$ and  stop iterations when $|f(\alpha)|<\epsilon^{\star}$ where $\epsilon^{\star}\leq0.001$
in all our results achieving accuracy of $1\times10^{-5}$ for a unit
premium.

In the continuous-time model \citet{RefWorks:7} use numerical PDE techniques to solve for $V_{0}$, 
corresponding to \citet[equation.~(7)]{Hyndman-Wenger-Decomp}, and present the fair fees for various
$(g,\sigma)$ combinations. In \citet{RefWorks:43}, a discrete-time
model is developed and the contract values are estimated using Monte Carlo simulation with a geometric
mean strike Asian call option as a control variate. Both papers assume
$S$ is log-normally distributed. In theory we expect convergence of
results for both models and our binomial model. However \citet{RefWorks:43}
obtains results significantly lower than those of \citet{RefWorks:7},
and concludes that the results of \citet{RefWorks:7}  are
on average 28\% too high.

\prettyref{tab:V0comparresuts} provides a comparison between
the results of \citet{RefWorks:7}, \citet{RefWorks:43}, and the binomial model. 
In the discrete models $\delta t=1/\text{time-steps}$.
The parameters are: $P=100$, $g=10\%$, $r=5\%$, $\sigma=20\%$,
$T=1/g=10$. For $\delta t=1$, results from the binomial model and
\citet{RefWorks:43} are sufficiently close. We reach three time-steps
per year under Method B, and observe that the binomial model supports the 
results of \citet{RefWorks:43}.
\begin{table}
\begin{centering}
\scalebox{.9}{
\begin{tabular}{lccccccc} \toprule
& M\&S (2006) &\multicolumn{3}{c}{\citet{RefWorks:43}} &\multicolumn{3}{c}{Binomial}\\ \cmidrule{2-8}
time-steps/year & continuous & $1$ & $12$ & $4000$ & $1$ & $2$ & $3$\\ \midrule
$\alpha^\star$(bps) & $140$ & $92.41$ & $96.65$ & $97.28$ & $92.20$ & $94.55$ & $95.35$\\ \bottomrule
\end{tabular}}
\par\end{centering}
\caption[Comparison of fair fee with previous results - no lapses]{\label{tab:V0comparresuts}Comparison of results for $\alpha^{\star}:\ g=10\%,\ r=5\%,\ \sigma=20\%$}
\end{table}

For the same parameters \prettyref{tab:Computational-Time-Comparison}
displays sample run-times (in seconds) to calculate $V_{0}$ for a
single value of $\alpha$. The differences may seem small for $n<3$
and external factors also affect the run-times.  However Method A is implemented
in \textit{Matlab} while Method B is implemented in 
\CC\, which is generally more efficient for identical code. Therefore, we find
that Method B is significantly slower. Under Method B with $n=3$
and $\alpha=95.35$bps, we observe that $W_{N}=0$ for all paths with
less than 11 up-moves and, therefore, the bottom 10 nodes in the recombining
tree for $Z$ do not need to be evaluated. However, this 
simplification does not prevent the run-time from  growing rapidly with $n$.%
\begin{table}
\begin{centering}
\begin{tabular}{lrr} \toprule
Time-Steps & Method A & Method B \\
& (Trees, Matlab) & (Loop, \CC)\\ \midrule
$n=1$ & $7.7\times10^{-4}$ & $3\times10^{-3}$ \\
$n=2$ & $0.80$ & $2.5$ \\
$n=3$ & & $3\times10^3$\\ \bottomrule
\end{tabular}
\par\end{centering}

\caption[Computational time comparison]{\label{tab:Computational-Time-Comparison}Computational time comparison
(in seconds)}

\end{table}

While the binomial model is a valuable theoretical framework for viewing
the GMWB rider, it is the Asian approximation method which reveals
the practical value of such a model. Implementing the Asian approximation
method, we attain results up to $n=10$. Monthly time-steps should
be attainable with more efficient programming and superior hardware.
The results in \prettyref{tab:ApproxMethods}  imply convergence
to the $\faira$ computed by \citet{RefWorks:43}.%
\begin{table}
\begin{centering}
\scalebox{0.9}{
\begin{tabular}{lccccccc} \toprule
 n & 1 & 2 & 3 & 5 & 7 & 9 & 10\\  \midrule
$\alpha^\star$(bps) & $92.30$ & $94.64$ & $95.40$ & $96.05$ & $96.33$ & $96.48$ & $96.54$\\
$V_0(\alpha=97.3)(\$)$ & 99.767 & 99.880 & 99.917 & 99.945 & 99.958 & 99.965 & 99.967\\
\bottomrule
\end{tabular}}
\par\end{centering}
\caption{\label{tab:ApproxMethods}Asian approximation results}

\end{table}

\prettyref{tab:alphaResutls-1} contains additional results for different
$g$ and $\sigma$ values. The fair fee is increasing with both $g$
and $\sigma$ and is quite sensitive to $\sigma$. Sensitivity results
have been discussed at length in the literature (see \citet{RefWorks:18}).
The return of premium guaranteed by the GMWB does not include time
value of money and as $g$ increases, the maturity decreases and $V_{0}$
increases in value for any fixed $\alpha$ because of the interest
rate effect. Consequently $\faira$ must increase. Our results consistently
support \citet{RefWorks:43} at the expense of \citet{RefWorks:7}.%
\begin{table}
\begin{centering}
\begin{threeparttable}[b]
\begin{tabular}{cccccccc} \toprule
\multicolumn{2}{l}{$(\faira$, bps)} & \multicolumn{3}{c}{$\sigma=20\%$} & \multicolumn{3}{c}{$\sigma=30\%$}\\\cmidrule(r){3-5} \cmidrule(l){6-8}
$g\%$ & $T$ & MS\tnote{a} & L\tnote{b} & B\tnote{c} & MS\tnote{a} & L\tnote{b} & B\tnote{c} \\ \midrule
5 & 20& 37 & 28.5 & 27.1(1) & 90  & 76.5  & 74.8(1)\\
6& 16.67& 54 &40.6 & 38.7(1) & 123  & 103.7  & 101.5(1)\\
7& 14.29& 73 &53.8 & 51.3(1) & 158 & 132.3  & 129.4(1)\\
8& 12.5& 94 & n/a & 64.6(1) & 194  & n/a & 158.3(1)\\
9& 11.11& 117 &n/a & 80.1(2) & 232  &n/a  & 189.3(2) \\
10& 10& 140 &96.7 & 94.6(2) & 271 & 221.2 & 219.1(2) \\\bottomrule
\end{tabular}
\begin{tablenotes}[para,flushleft]
\begin{footnotesize}
\item[a] \citet{RefWorks:7}
\item[b] \citet{RefWorks:43} with $n=12$
\item[c] Binomial with $n$ in parentheses
\end{footnotesize}
\end{tablenotes}
\end{threeparttable}
\par\end{centering}

\caption[Comparison of fair fee with previous results - no lapses]{\label{tab:alphaResutls-1}Comparison with previous results for $\alpha^{\star}$,
$(r=5\%)$}

\end{table}

\begin{figure}
\begin{centering}
\scalebox{0.75}{
   \includegraphics{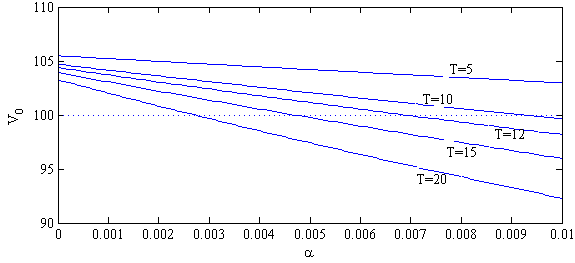}}
\par\end{centering}

\caption[$V_{0}$ as a function of $\alpha$ for varying $T$ - no lapses]{\label{fig:Vsens}Plotting $V_{0}$ as a function of $\alpha$ for
varying $T$. Parameters are: $r=5\%$, $\sigma=20\%$, and $g=1/T$.}

\end{figure}

In \prettyref{fig:Vsens}, $V_{0}$ is plotted against $\alpha$ for
different $T$ values. The parameters are: $P=100$, $r=5\%$, $\sigma=20\%$,
$\delta t=1$, and $g=(1/T)$. The fair fee is the point of
intersection between the horizontal line $V_{0}=100$ and the curves.
When the curves are plotted over the wider range {[}0,0.05{]} the
linearity resemblance seen on $[0,0.01]$ disappears and the curves
have a more pronounced convex shape. As $\alpha$ increases, the likelihood
of trigger rises but the decrease in the expected discounted terminal
account value is less sensitive for sufficiently large $\alpha$. 

It is important to consider the sensitivity of $V_{0}$ to $\alpha$
in a neighbourhood around $\faira$, for a given set of parameters.
\prettyref{fig:Vsens} reflects the changing sensitivity for different
values of $T$. For the parameters in \prettyref{tab:V0comparresuts},
the binomial method with $\delta t=2$ gives $V_{0}(100,140\text{ bps},10\%)=98.02$
and it can be deceptive to only look at $\faira$. The objective is
to solve for the fair fee and in our pricing framework, charging a
different fee leads to arbitrage no matter the size of $|\alpha-\faira|$.
However, in the presence of  real world constraints such as imperfect models, market
frictions,  and  sub-rational policyholder behaviour small pricing 
errors may not lead to arbitrage and it is crucial consider price
sensitivity in addition to finding $\faira$.%

\subsection{Distribution of the Trigger}

\citet{RefWorks:7} numerically solve the Kolmogorov backward equation
for $\mathbb{P}(\tau\leq T)$ and provide results for different combinations
of $(\mu,\sigma)$ with the parameters $g=7\%$ and $\alpha=40$bps.
To avoid fractional years,
we set $T=14$ and $g=7.14\%$. As shown in \prettyref{tab:-comparing-binomial}, the binomial model with just $n=2$ produces probabilities close to  \citet{RefWorks:7}. The accuracy improves with increasing $\sigma$.  

In \citet{RefWorks:7}, $S_{t}$ is modelled by geometric Brownian
motion

\[
dS_{t}=\mu S_{t}dt+\sigma S_{t}dB_{t}^{'},\] 

\noindent
where $B_{t}^{'}$ is $\mathbb{P}$-Brownian motion. Then with

\[
r_{T}^{s}:=\ln\left(\frac{S_{T}}{S_{0}}\right)=\left(\mu-\frac{1}{2}\sigma^{2}\right)T+\sigma B_{T}^{'},\] 

\noindent
we have $\mathbb{E_{P}}[r_{T}^{s}]=(\mu-\frac{1}{2}\sigma^{2})T$
and $Var_{\mathbb{P}}[r_{T}^{s}]=\sigma^{2}T$. In the binomial model we set

\begin{align*}
u & =e^{\sigma\sqrt{\delta t}},\\
d & =e^{-\sigma\sqrt{\delta t}},\\
\tilde{p} & =\frac{1}{2}+\frac{1}{2}\left(\mu-\frac{1}{2}\sigma^{2}\right)\frac{1}{\sigma}\sqrt{\delta t}.\end{align*}

\noindent
Note that $\tilde{p}<1$ holds only if $\mu<\frac{1}{2}\sigma^{2}+\sigma\frac{1}{\sqrt{\delta t}}$.
For $\delta t=1$ this condition is violated for $\sigma=10\%$ and
$\mu=12\%$.

In general, the probability mass function of $\tau$ with respect to $\mathbb{P}$
can be calculated in the binomial model using equation~\prettyref{eq:Markov prop Qtau-1},
where

\[
\mathbb{P}(\tau=i)=H(0,0,i,P)\] 

\noindent
for $i\in\{1,2,\dots,N,\infty\}$. Of course, $p$ must be replaced
with $\tilde{p}$.

\begin{table}
\begin{centering}
\scalebox{0.62}{
\begin{tabular}{*{13}{c}} %
 & \multicolumn{3}{c}{$\sigma=10\%$} & \multicolumn{3}{c}{$\sigma=15\%$} & \multicolumn{3}{c}{$\sigma=18\%$} & \multicolumn{3}{c}{$\sigma=25\%$}\\ \cmidrule{2-13}
 & M\&S & \multicolumn{2}{c}{Binomial} & M\&S & \multicolumn{2}{c}{Binomial}& M\&S & \multicolumn{2}{c}{Binomial}& M\&S & \multicolumn{2}{c}{Binomial}\\ \cmidrule{2-13}
 & & $\delta t=0.5$ & $\delta t=1$ & & $\delta t=0.5$ & $\delta t=1$& & $\delta t=0.5$ & $\delta t=1$& & $\delta t=0.5$ & $\delta t=1$\\ \cmidrule{2-13}
$\mu=4\%$ & 19.0\% & 16.0\% & 15.2\% & 31.4\% & 31.1\% & 30.9\% & 37.8\% & 38.2\% & 38.2\% & 49.9\% & 50.8\% & 50.6\%\\
$\mu=6\%$ & 7.0\% & 4.5\% & 3.6\% & 18.5\% & 17.8\% & 16.9\% & 25.5\% & 25.3\% & 25.0\% & 39.6\% & 40.5\% & 40.2\%\\
$\mu=8\%$ & 1.7\% & 0.7\% & 0.3\% & 9.3\% & 8.2\% & 7.4\% & 15.5\% & 15.0\% & 14.5\% & 30.5\% & 30.8\% & 30.4\%\\
$\mu=10\%$ & 0.3\% & 0.0\% & 0.0\% & 4.1\% & 3.1\% & 2.3\% & 8.6\% & 7.8\% & 7.1\% & 22.2\% & 22.2\% & 21.7\%\\
$\mu=12\%$ & 0.04\% & 0.0\% & - &1.6\% & 0.9\% & 0.4\% & 4.4\% & 3.5\% & 2.7\% & 15.5\% & 15.2\% & 14.4\%\\ \bottomrule
\end{tabular}}
\par
\end{centering}

\caption[Comparing finite trigger time probabilities with previous results]{$\mathbb{P}(\tau<\infty)$: comparing binomial mod\label{tab:-comparing-binomial}el
to continuous time model from \citet{RefWorks:7}}
\end{table}

\begin{Remark}
Applying equation~\prettyref{eq:Markov prop Qtau-1} to calculate the
trigger probabilities with two time-steps a year,
$2^{28}$ paths need to be evaluated and we run into capacity issues.
 For $\delta t=0.5$, we use the approach
of equation~\prettyref{eq:Vonovector} except that rather than working with
$\exp{(-rT)}W_{T}$, we use the indicator function $\mathbf{1}_{\{W_{T}=0\}}$
remembering to take account of the probabilities for the lower nodes
with more than $A_{0}$ down movements.
\end{Remark}

\subsection{Comparison of Hedging and No Hedging}

\begin{figure}
\begin{centering}
\scalebox{0.75}{
   \includegraphics{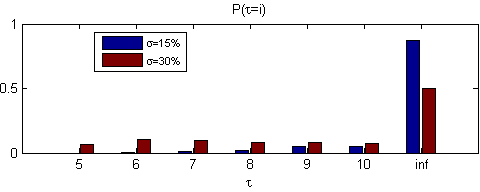}
   }
\caption[Probability mass function: effect of $\sigma$]{\label{fig:Probability-mass-function}Probability mass function of $\tau$: different volatilities
}
\end{centering}
\end{figure}

We investigate the impact of volatility on the fees, triggers and
losses. The parameters are: $g=10\%,\ T=10,\ P=100,$ and $\delta t=1$.
The risk free rate $r$ is $5\%$ and the drift term $\mu$ of the
underlying asset is $7.5\%$. We consider $\sigma=15\%$ and $\sigma=30\%$.
The respective fair fees $\faira$ are $41.8$bps and $216.7$bps.
The probability mass function for $\tau$ under the physical measure
is displayed in \prettyref{fig:Probability-mass-function}. Recall
that $\tau=\infty$ when $W_{T}>0$. The two $\sigma$ values were
selected to magnify the interaction between volatility, the trigger
time distribution and consequently the rider payouts. Higher volatility
implies more adverse market returns and a greater likelihood of early
trigger. An additional effect on trigger comes from the rider fee.
The fee rate is very sensitive to volatility and the fees drag down
the account value further, resulting in more frequent early trigger
times. 

\begin{figure}
\subfloat[]{
   \includegraphics[scale=.75]{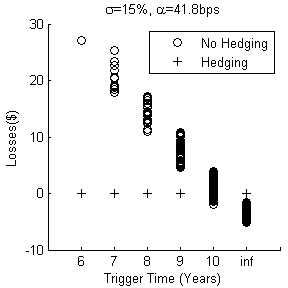}
}
\subfloat[]{
  \includegraphics[scale=0.75]{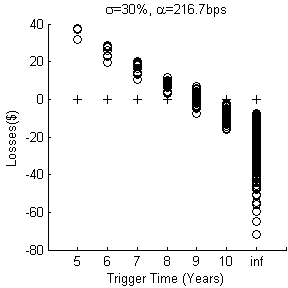}
}
\caption[Hedging and no-hedging losses (no lapses)]{\label{fig:HedgeNoHedge} Hedging and no-hedging losses, with $r=5\%$
and $g=10\%$}
\end{figure}

We consider the strategies of no hedging and dynamic delta hedging
prescribed in \prettyref{sub:Hedging}. Define $\Pi:=\exp{(-\barr N)}X_{N}$
to be the discounted profit. When $\Delta$ follows the prescribed
portfolio process \prettyref{eq:DeltaBinNolapse} we obtain the hedging
profit, $\Pi^{H}$. If $\Delta\equiv0$ we obtain the profit under
no hedging, $\Pi^{NH}$. The superscripts are omitted when it is clear
which profits we are analyzing. \prettyref{fig:HedgeNoHedge} plots
both $-\Pi^{H}$ and $-\Pi^{NH}$ against $\tau_{0}$ for the complete
set of outcomes ($2^{10}=1024$ paths). The values are per \$100 initial
premium.

\begin{figure}
\begin{centering}
\includegraphics[scale=0.9]{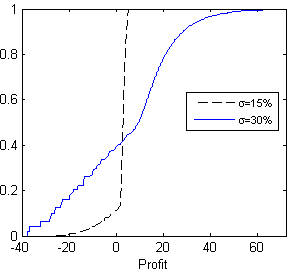}\caption[Cumulative distribution function of profit with no hedging]{\label{fig:CDF-of-profits}CDF of $\Pi^{NH}$ with respect to $\mathbb{P}$}
\end{centering}
\end{figure}

The dynamic delta hedging strategy results in no losses. Without hedging,
the range of potential losses by each random trigger time has a decreasing
trend because a later trigger time implies additional periods of fee
revenue and fewer periods of any rider guarantee payout. The effect
of the volatility $\sigma$ is particularly visible for those pathwise
outcomes where $\tau=\infty$. When $\sigma=15\%$ there is an $87\%$
probability of a positive terminal account value but the gains are
small. On the other hand, there is only a $50\%$ probability that
$\tau=\infty$ when $\sigma=30\%$ but the potential profits are large
due to the high fees. \prettyref{fig:CDF-of-profits} shows the cumulative
distribution function of the profits when there is no hedging.

We present several risk measures for the no-hedging profit $\Pi^{NH}$ under $\mathbb{P}$. The
standard deviation is denoted $SD(\Pi)$. The tail value at risk is
$TVaR_{\gamma}(\Pi):=E_{\mathbb{P}}[-\Pi|\Pi\leq-VaR_{\gamma}(\Pi)]$
where $VaR_{\gamma}(\Pi)=-\inf\{x:\mathbb{P}(\Pi\leq x)>\gamma\}$.
\prettyref{tab:Statistics1} shows the values for this sensitivity
analysis of $\sigma$. Using the real world probability measure 
only amplifies the effect of $\sigma$ on
the insurer's risk and highlights the importance of a thorough hedging
scheme.%

\begin{table}
\begin{centering}
\begin{tabular}{lcc} \toprule
Values per \$100 & $\sigma=15\%$ & $\sigma=30\%$ \\ \midrule
$E_\mathbb{P} (\Pi^{NH})$ & 1.84 & 4.19\\
$\text{SD}_\mathbb{P} (\Pi^{NH})$ & 4.28  & 21.34\\
$TVaR_{0.10} (\Pi^{NH})$ & 9.30 & 32.60 \\\bottomrule
\end{tabular}
\par\end{centering}

\caption{\label{tab:Statistics1}Profit metrics for no hedging (no lapses)}

\end{table}

\subsubsection{Hedging in a Continuous Model}

In the binomial model a perfect hedge is attainable. Suppose instead
the underlying asset follows the geometric Brownian motion process
given by \prettyref{eq:StP}. A perfect hedge in this case entails
continuously re-balancing the hedging positions by taking a position
at any time t of $\frac{W}{S}\frac{\partial U}{\partial W}$ units
of $S$ (see \citealp{RefWorks:18}). In practice, the positions will
be rebalanced only a finite number of times each year which introduces
hedging errors. We model the fees and withdrawals to occur only at
year-end in order to contrast with the previous result in the binomial
model for $\delta t=1$. This differs from the continuous model of
\citet{Hyndman-Wenger-Decomp} where fees and withdrawals
are deducted continuously.

\begin{figure}
\begin{raggedright}
\subfloat[No hedging]{\includegraphics[scale=0.75]{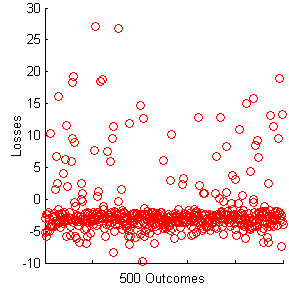}
}\subfloat[Weekly hedging]{\includegraphics[scale=0.75]{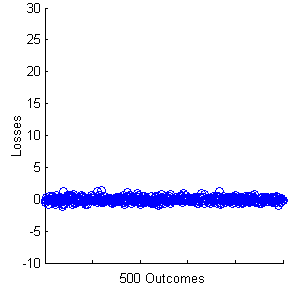}
}
\par\end{raggedright}
\caption[Weekly hedging versus no hedging in continuous model]{\label{fig:50-simulated-outcomes}Continuous model with $g=$10\%,
$r=$5\%, $\mu=7.5\%$ , $\sigma$=15\%, $\alpha$=45bps}
\end{figure}

The parameters used are $P=100,\ g=10\%,\ r=5\%,\ \mu=7.5\%,\ \sigma=15\%$,
and $T=10$. We used Monte Carlo simulation to obtain $\alpha^{\star}\approx45$bps
(50,000 paths were simulated).
We analyzed the effectiveness of a dynamic hedging strategy with weekly
re-balancing for 500 path outcomes generated under $\mathbb{P}$.
For $t\in\{0,\frac{1}{52},\frac{2}{52},\dots,\frac{519}{52},10\}$
and $w\in\mathbb{R}_{+}$, Monte Carlo simulations (using 1000 paths)
yielded $U_{t}(w-1)$ and $U_{t}(w+1)$. We approximated $\frac{\partial U}{\partial W}$
with $\Delta_{t}(W_{t})={(U_{t}(W_{t}+1)-U_{t}(W_{t}-1))}/{2}$
where the same set of generated paths was used to obtain both values
in the numerator. Using the same paths and taking the central difference
has been shown to reduce variability of results (\citealp{RefWorks:36}).
\prettyref{fig:50-simulated-outcomes} displays the discounted losses
for no hedging and for weekly hedging for each generated path. Based
on the simulations, $\mathbb{P}(\tau=\infty)=84.4\%$. As supported
by \prettyref{tab:mean,std,discretehedgerrors}, the weekly hedging
considerably mitigates the equity risk.  
In contrast to the case when the underlying model is binomial, negative hedging
errors arise when the underlying model is continuous.
 \begin{table}
\begin{centering}
\begin{tabular}{ccc} \toprule
Values per \$100 & No Hedging & Hedging (Weekly) \\ \midrule
$E_\mathbb{P} [\Pi]$ & 1.86 & 0.07\\
$\text{SD}_\mathbb{P} [\Pi]$ & 4.63  & 0.36\\
$TVaR_{0.10} (\Pi)$ & 10.15 & 0.61 \\\bottomrule
\end{tabular}
\par\end{centering}
\caption{\label{tab:mean,std,discretehedgerrors}Profit metrics for continuous
model with weekly hedging and no hedging}
\end{table}

\subsection{The Fair Rider Fee with Surrenders}

We next compare our results for $\faira$ when early surrenders
are permitted with those in the literature.
For the parameter set of $g=7\%,$ $r=5\%$, and $k_{i}=1\%$ for
all $i$, \prettyref{tab:alphaResutlsLapses} compares the binomial
model with $\delta t=1$ to \citet{RefWorks:7}. Although the results
are proportionally closer, as compared to \prettyref{tab:V0comparresuts},
it is inconclusive if the differences are mostly due to $\delta t=1$
or if the results presented by \citet{RefWorks:7} in the lapse case
suffer from the same inaccuracies as in the no-lapse case.%
\begin{table}
\begin{centering}
\begin{tabular}{lccccc} \toprule
$\sigma(\%)$ & 15 & 18 & 20 & 25 & 30 \\ \midrule
\citet{RefWorks:7} & 97 & 136& 160 & 320 & 565 \\
Binomial ($\delta t=1$) & 33 & 89 &138 & 283 & 455\\\bottomrule
\end{tabular}
\par\end{centering}

\caption[Comparison of fair fee with \citet{RefWorks:7} (lapses)]{\label{tab:alphaResutlsLapses}Comparison of $\alpha^{\star}$ to
previous results; with $g=7\%$, $r=5\%$, and $k=1\%$.}

\end{table}

We apply the Asian approximation method with the parameters $g=10\%$,
$r=5\%$, $\sigma=20\%$, and $k=3\%$ in \prettyref{tab:ApproxMethods_Lapses}.
The convergence is slower than in the no-lapse case, but that is a
result of the early surrender decisions which are being approximated.
This is consistent with the findings of \citet{RefWorks:84}. The
rightmost column shows $\faira$ under the original binomial model.
The increase in $\faira$ when $n$ is increased from one to two suggests
that a sizable portion of the differences in \prettyref{tab:alphaResutlsLapses}
can be attributed to the low value of $n$ in the binomial model.%
\begin{table}
\begin{centering}
{\setlength{\tabcolsep}{.3em}
\begin{tabular}{lccc} \toprule
 n & $\alpha^\star$(bps) & $V_0(\alpha\negmedspace=\negmedspace 146.4)(\$)$ & $\alpha^\star$(actual)\\  \midrule
 1 & $131.00$ & 99.689 & 130.54\\
2& $141.98$ & 99.933 & 141.75\\
3& $143.37$ &  99.949\\
4& $146.04$ & 99.994\\
5& $146.40$ & 100\\
6& $146.70$ & 100.005\\ \bottomrule
\end{tabular}}
\par\end{centering}

\caption{\label{tab:ApproxMethods_Lapses}Asian approximation results - lapses}

\end{table}

We set $r$ equal to the instantaneous risk-free rate long term mean
and $\sigma$ equal to the variance long term mean used in the stochastic
interest rate and volatility processes in \citet{RefWorks:61}. We
found that comparing $V_{0}$ for varying $\alpha$, in the no lapse
case the binomial model provides close estimates even for $\delta t=0.5$.
In \prettyref{tab:alphaResutlsLapses_Bac} we list the difference
in the contract value between the two methods for varying $\alpha$
and $P=100$, $g=10\%$, $r=3\%$, $\sigma=20\%$, and $k=3\%$. The
models have fundamental differences and we do not expect to attain
exact results in the limit.%
\begin{table}
\begin{centering}
\begin{threeparttable}[b]
\begin{tabular}{lccccc} \toprule
$\alpha(\%)$ & 1 & 2 & 3 & 4 & 5\\ \cmidrule(l){2-6}
$V_0^B(\alpha)-V_0^{BMOP}(\alpha)^{a,b}$: (no lapse) & -0.186 & -0.113 & -0.035 & 0.05 & 0.096\\ 
$V_0^B(\alpha)-V_0^{BMOP}(\alpha)$: (lapse) & 0.153 & 0.546 & 0.75 & 0.78 & 1.04 \\\bottomrule
\end{tabular}
\begin{tablenotes}[para,flushleft]
\begin{footnotesize}
\item[a] $V_0^B$ refers to the binomial method, with $\delta t=0.5$.\\
\item[b] $V_0^{BMOP}$ refers to \citet{RefWorks:61}.
\end{footnotesize}
\end{tablenotes}
\end{threeparttable}
\par\end{centering}

\caption[Comparison of contract values with \citet{RefWorks:61}]{\label{tab:alphaResutlsLapses_Bac}Comparison of $V_{0}$ with previous
results: $g=10\%$, $P=100$, $r=3\%,\ \sigma=20\%,$ and $k=3\%$.}

\end{table}

Sensitivity results for $g,\ r,$ and $\sigma$ are shown in \prettyref{tab:alphaResutlsLapse}.
The baseline case is set to $g=10\%,\ r=5\%,\ \sigma=20\%,$ and a
CDSC of $k=3\%$. The fair fee $\faira$ is increasing with $g$ and
$\sigma$ but decreasing with $r$, however, the fair fee is most sensitive to $r$.
The sensitivity of the fair fee to $r$ is due to the long duration of the contract. 
Therefore, incorporating a stochastic interest rate model is justified, though beyond the
scope of this paper.%
\begin{table}
\begin{centering}
\begin{threeparttable}[b]
\begin{tabular}{ccccccccc} \toprule
$g\%$ & $\faira$ (bps) & $V_0 (\alpha_1)$ (\$) & $\sigma\%$ & $\faira$ & $V_0 (\alpha_1)$ & $r\%$ & $\faira$ & $V_0 (\alpha_1)$ \\
\cmidrule(r){1-1} \cmidrule(lr){2-3} \cmidrule(r){4-4} \cmidrule(lr){5-6} \cmidrule(r){7-7} \cmidrule(lr){8-9}
5 & 30& 97.21 & 10 & 10 & 97  & 1  & 1199 & 108.21\\
6 & 47& 97.87 &15 & 44 & 97.84  & 2  & 673 & 105.54\\
7 & 68& 98.44 &18 & 87 & 99.08 & 3  & 397 & 103.29\\
8 & 90& 98.95 &20 & 142 & 100  & 4 & 244 & 101.43\\
9& 110& 99.38 &25 & 318 & 102.46  &5  & 142 & 100 \\
10& 142& 100 &30 & 562 & 105.12 & 6 & 77 & 98.87 \\\bottomrule
\end{tabular}
\begin{tablenotes}[para,flushleft]
\begin{footnotesize}
\item[a] Baseline case is $g=10\%,\,r=5\%,\,\sigma=20\%,\,k=3\%$, $\alpha_1 =142$bps.
\item[b] For the first column, $\delta t=1$ for $g\leq9\%$. All other values use $\delta t=2$.
\end{footnotesize}
\end{tablenotes}
\end{threeparttable}
\par\end{centering}

\caption[Sensitivity results for fair fee]{\label{tab:alphaResutlsLapse}Sensitivity results for $\alpha^{\star}$}

\end{table}

Under the parameters $g=10\%$, $r=5\%$, $\sigma=25\%$, and $\delta t=1$,
the impact of the CDSC schedule on $\alpha^{\star}$ is shown in \prettyref{tab:lapse k alpha}.
Allowing surrenders with no penalties, the fair fee will be exorbitant
to compensate for this option. As the penalties increase, the fee
approaches the corresponding fee in the no-lapse model. For sufficiently
high penalties, the option to surrender yields no marginal value.%
\begin{table}
\begin{centering}
\begin{tabular}{lc}\toprule
Description of Schedule & $\alpha^\star$(bps)\\\midrule
No-Lapse Model & \bf{152}\\
$k_i =0$ for $i=1,\dots,9$ & 491\\
$k_i =1\%$ for $i=1,\dots,9$ & 430\\
$k_i =3\%$ for $i=1,\dots,9$ & 309\\
$k_i =5\%$ for $i=1,\dots,9$ & 217\\
$k_i =7\%$ for $i=1,\dots,9$ & 169\\
$k_i =8\%$ for $i=1,\dots,9$ & 155\\
$k_i\geq 8.38 \%$ for $i=1,\dots,9$ & \bf{152}\\
$k_i =(10-i)\%$, for $i=1,\dots,9$ & 171\\
$k_i =(9-i)\%$, for $i=1,\dots,9$ & 188\\\bottomrule
\end{tabular}
\par\end{centering}

\caption[Impact of surrender charges on fair fee]{\label{tab:lapse k alpha}Impact of $k$ on $\alpha^{\star}$}

\end{table}

\subsection{Hedging and No Hedging with Surrenders}

\begin{figure}
\begin{centering}
\subfloat[\label{fig:Nolapses}No lapses]{\includegraphics[scale=0.75]{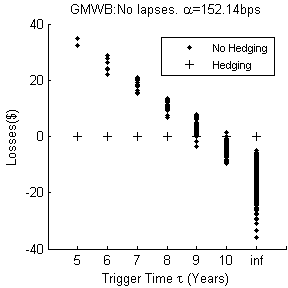}

}\subfloat[\label{fig:Lapses_Hedging}SC begins at 8\% and decreases 1\% per
annum]{\includegraphics[scale=0.75]{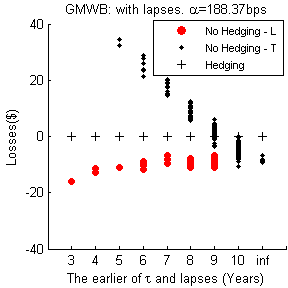}

}
\par\end{centering}
\centering{}\caption[Hedging and no hedging, with and without lapses]{Hedging and no hedging, with and without lapses: $g=10\%$, $r=5\%$,
$\sigma=25\%$.\label{fig:Comparing LapseNoLapse} }

\end{figure}

We consider the parameters: $P=100$, $g=10\%$, $r=5\%$, $\sigma=25\%$,
and $\delta t=1$. The drift of $S$ is $\mu=7.5\%$. The surrender
charge schedule applied is $k_{i}=\max(.09-.01i,0)$ for $i=1\dots10$.
\prettyref{fig:Comparing LapseNoLapse} plots the aggregate losses,
discounted to time zero, for the set of all outcomes for both the
no-surrender model and the model with early surrenders. The respective
fair fees are charged. In \prettyref{fig:Lapses_Hedging} the no-hedging
results are denoted by L and T: the former are outcomes where it is
optimal to lapse while the latter are those for which no lapse occurs.

\prettyref{tab:Pmf_CompareLapseNoLapse} shows the $\mathbb{P}-$distribution
of trigger times and surrender times, where $\eta^{\star}$ denotes
an optimal early surrender. Note that $\mathbb{P}(\tau=\infty)\approx60\%$
when surrenders are not allowed, but this reduces to $\mathbb{P}(\tau=\infty)\approx0.65\%$
when surrenders are permitted. Allowing lapses causes a shift as it
becomes preferable in many outcomes when the market is doing well
for the policyholder to lapse rather than face the likelihood of the
rider maturing without being triggered.

For the outcomes where it is optimal to lapse, the profits to the
insurer are decreasing for years 3 to 7. This is due to the design of
the surrender charge schedule $k_i$. The higher surrender charge in earlier years outweighs the additional
fees received when lapses occur later.%
\begin{table}
\begin{centering}
\begin{tabular}{cccc} \toprule
 & No Lapses & \multicolumn{2}{c}{Model with Lapses}\\ \cmidrule{2-4}
$i$ & $\mathbb{P}(\tau=i)$ & $\mathbb{P}(\tau=i)$ & $\mathbb{P}(\eta ^\star =i)$\\ \midrule
3 & 0 & 0\% & 20.28\%\\
4& 0 & 0\% &16.73\%\\
5& 2.90\% &2.90\%&4.91\%\\
6&5.80\% &5.80\%&8.11\%\\
7&7.83\% &7.83\%&3.57\%\\
8&6.29\% &9.08\%&4.42\%\\
9&9.48\% &7.37\%&2.37\%\\
10&7.23\% &5.98\%&0\\
$\infty$ & 60.47\% & .65\%& 0\\\midrule
Sum & 1 & 39.61\% & 60.39\%\\ \bottomrule
\end{tabular}
\par\end{centering}

\caption{Probability dis\label{tab:Pmf_CompareLapseNoLapse}tribution of $\tau$
and lapses for \prettyref{fig:Comparing LapseNoLapse}}

\end{table}

\begin{figure}
\centering{}\includegraphics{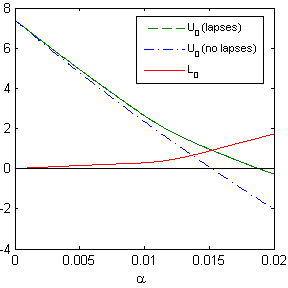}\caption[Value of option to lapse]{\label{fig:ValueLapse}Value of $L_{0}$: $g=10\%,$ $r=5\%$, $\sigma=25\%$,
$\delta t=1$, and a declining SC schedule }
\end{figure}

Numerical results for the value of the option to surrender, $L_{0}$, are presented in \prettyref{fig:ValueLapse}. 
When $\alpha$  is small, there is little incentive to surrender early and $L_{0}\approx0$.
For larger values of $\alpha$ there is incentive to surrender and avoid paying future fees. This relationship
is reflected in the growth of $L_{0}$ with $\alpha$.%

\section{\label{sec:bin mort}Extending the Model: Including Mortality Risk}

The simplification of disregarding mortality was used in several papers
for GMWBs including \citet{RefWorks:7} and \citet{RefWorks:8}. Mortality
factors do need to be considered in practice. Depending on the goal
of the analysis, the level of precision attained by including mortality
may not justify the added complexity and dimensionality of the model.
In particular, in the papers mentioned the focus was on studying the
optimal policyholder behaviour strategy and including mortality only
detracts from the presentation of the results.

Mortality risk is typically assumed to be independent of financial
risk. Further, under the assumption of independent lives and deterministic
forces of mortality (hazard rates) a simple application of the strong
law of large numbers justifies the claim that mortality risk is diversifiable.
By issuing a sufficiently large portfolio of homogeneous policies
the insurer can completely account for the mortality risk by taking
the expected value of claim payments under the appropriate mortality
probability distribution (\citealp{RefWorks:38}). Therefore under
these assumptions mortality risk is not priced by capital markets
in an economic equilibrium (no-arbitrage) approach and there is no
difference between the physical and risk-neutral measures (\citealp{RefWorks:60}).
In a stochastic mortality framework the non-diversifiable component
of mortality risk must be priced into the contract. 

\citet{RefWorks:60} list capacity constraints in immediate annuity
markets as one of several industry trends which justify charging for
mortality risk. We remark that in variable annuity markets, both finite
demand and regulatory limits on capital at risk lend support to modeling
capacity constraints in order to determine whether there is a non-negligible
impact.

The effect of mortality for GMWBs clearly depends on the death benefits
(DBs). When benefit payments are similar for both death and survival,
there is minimal impact. Indeed, \citet{RefWorks:61} found that guaranteed
minimum death benefit (GMDB) riders add little value to the contract
in the presence of other living benefit riders and a relatively short
maturity. 

We extend the model from \prettyref{sec:Bin wo mort} and \prettyref{sec:Binlapses} to include mortality
under the independence of lives assumption and deterministic forces
of mortality. It is straightforward to obtain the price processes
$V$ and $U$, which for each insured are dependent on the survival
status. The rider fee is obtained assuming diversifiable mortality
risk, as is the hedging portfolio; however, we consider a numerical
simulation to emphasize that under capacity constraints and finite
number of policies there is mortality risk and the product is not
fully hedged.

\subsection{\label{sec:Mortality-Risk-Model}Mortality Risk Framework}

In this section we establish a mortality framework. The classical
actuarial theory and notation used follows that of \citet{RefWorks:39}.
In addition, the measure-theoretic aspects and inclusion of counting
processes follows closely the frameworks of \citet{RefWorks:81} and
\citet{RefWorks:45}. 
\begin{Assumption}
\label{ass:iid}Homogeneous policies are issued to a pool of $l_{x}$
policyholders, each of age $x$. Measured from issue date, the random
times of death, denoted by $\{T_{j}^{x};j=1,\dots,l_{x}\}$ where
$T_{j}^{x}$ is the time of death for policyholder $j$, are absolutely
continuous,  independent and identically distributed, and lie on
a probability space $(\Omega^{M},\mathcal{F}^{M},\mathbb{P}^{M})$. 
\end{Assumption}
Consider a representative random variable $T^{x}$ where $T^{x}$ has the same distribution as $T_{j}^{x}$.
The support of $T^{x}$ is $[0,T^{\star})$ where $T^{\star}\leq\infty$
is the maximum remaining lifetime for a person age $x$. Corresponding
to the binomial model with $\delta t=1/n$ and $n\in\mathbb{N}_{+}$,
let $K^{x}$ denote the period in which death occurs. Then $K^{x}=\lceil T^{x}/\delta t\rceil$.
In other words, $K^{x}=i$ is equivalent to $(i-1)\delta t<T^{x}\leq i\delta t$.
For $j=1,\dots,l_{x}$, define the counting processes

\[
D^{x,j}=\{D_{i}^{x,j}:=\mathbf{1}_{\{K_{j}^{x}\leq i\}};i=1,\dots,N\}.\] 

\noindent
We work with the filtration generated by $\{D^{x,j}\}_{1\leq j\leq l_{x}}$.
The filtration is $\mathbb{F}^{M,x}:=\{\mathcal{F}_{i}^{M,\{x,l_{x}\}}\}_{1\leq i\leq N}$
where $\mathcal{F}_{i}^{M,\{x,l_{x}\}}:=\mathcal{F}_{i}^{M,x,1}\vee\dots\vee\mathcal{F}_{i}^{M,x,l_{x}}$
and $\mathcal{F}_{i}^{M,x,j}=\sigma(D_{l}^{x,j};l=1,\dots,i).$ We
work with the resulting filtered probability space $(\Omega^{M},\mathcal{F}_{N}^{M,\{x,l_{x}\}},\mathbb{F}^{M,x},\mathbb{P}^{M})$.
\begin{Remark}
The notation $\mathcal{G}\vee\mathcal{H}$, where $\mathcal{G}$ and
$\mathcal{H}$ are $\sigma$-algebras, means the $\sigma$-algebra
generated by $\mathcal{G}\cup\mathcal{H}$.
\end{Remark}
We define the process which produces $1$ while the insured $j$ is
still alive by $A_{i}^{x,j}:=1-D_{i}^{x,j}$ for $i\in\mathcal{I}_{N}$.

By \prettyref{ass:iid}, $T^{x}$ has a density function $f_{T^{x}}$.
Its cumulative distribution function is denoted $F_{T^{x}}(t):=\mathbb{P}(T^{x}\leq t).$ The deterministic
force of mortality, $\mu_{x}(t)$, is defined as the conditional probability
density function of $T^{x}$ at time $t$, given survival to that
time. Then

\begin{equation}
\mu_{x}(t):=\frac{f_{T^{x}}(t)}{1-F_{T^{x}}(t)}.\label{eq:forcemort}
\end{equation}

\noindent
 We introduce some additional actuarial notation:

\begin{align*}
_{j}p_{x+i}: & =\mathbb{P}(K^{x}>i+j|K^{x}>i)=\mathbb{P}(T^{x}>(i+j)\delta t|T^{x}>i\delta t),\\
_{j|l}q_{x+i}: & =\mathbb{P}(i+j<K^{x}\leq i+j+l|K^{x}>i),
\end{align*}

\noindent
and we write $p_{x+i}$ for $_{1}p_{x+i}$, $_{j}q_{x+i}$ for $_{0|j}q_{x+i}$,
and $q_{x+i}$ for $_{1}q_{x+i}$. It follows that 

\begin{align*}
_{i}q_{x}=F_{T^{x}}(i\delta t), \quad & 
_{i}p_{x}=1-{}_{i}q_{x}, \
\mbox{and} \quad 
_{j|l}q_{x+i}={}_{j+l}q_{x+i}-{}_{j}q_{x+i}.
\end{align*}

\noindent
From \prettyref{eq:forcemort} we have $f_{T^{x}}(i\delta t)=\mu_{x}(i\delta t){}_{i\delta t}p_{x}$
and $_{j}p_{x+i}=e^{-\int_{0}^{j\delta t}\mu_{x+i\delta t}(u)du}$
(see \citet{RefWorks:39} for details). Note that $F_{T^{x}},f_{T^{x}},$
and $\mu_{x}$ are defined on the reals, while $_{j}p_{x+i}$ and
$_{j|l}q_{x+i}$ are defined on the integers.  \citet{RefWorks:39} provide several analytical laws of mortality.
\begin{Definition}
Under the Makeham law

\[
\mu_{x}(t):=A+Bc^{x+t}\] 

\noindent
where $B>0,\, A\geq-B,\, c>1$ and $x+t\geq0$.
\end{Definition}
As a result, under the Makeham law:

\[
_{i}p_{x}=\exp\left(-i\delta tA-\frac{B}{\ln(c)}(c^{x+i\delta t}-c^{x})\right).\] 

\noindent

\begin{Example}
\label{exa:The-parameters-used}The parameters used to develop the
illustrative life table under the Makeham law in \citet{RefWorks:39}
are: $A=0.7\times10^{-3},\ B=0.05\times10^{-3}\ \text{and }c=10^{0.04}$.
\prettyref{fig:probability-of-death} plots both $f_{T^{x}}(t)$ and
$\mathbb{P}(T^{x}>t)$ for $x=60$ and $t\in[0,50]$.%

\begin{figure}
\scalebox{0.9}{
\includegraphics{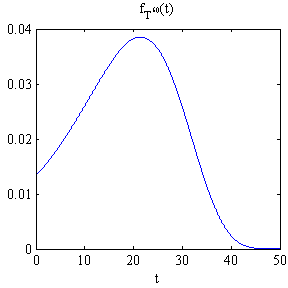}\includegraphics{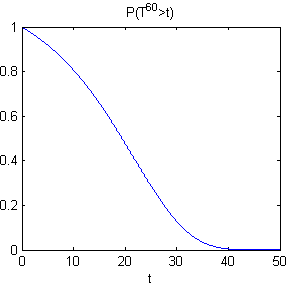}
}
\caption[Distribution of time of death, with Makeham law]{\label{fig:probability-of-death}Using the Makeham law, with $A=0.7\times10^{-3},\ B=0.05\times10^{-3}\ \text{and }c=10^{0.04}$ }

\end{figure}

\end{Example}
We state one additional useful result from \citet{RefWorks:45}. For
$i\leq j$,

\[
\mathbb{P}(T^{x}>j\delta t|\mathcal{F}_{i}^{M,x})=(1-D_{i}^{x}){}_{j-i}p_{x+i}\] 

\noindent
and

\[
\mathbb{P}(i\delta t<T^{x}\leq j\delta t|\mathcal{F}_{i}^{M,x})=(1-D_{i}^{x}){}_{j-i}q_{x+i}.\] 

\noindent

\subsection{Death Benefit Design}

We consider both the ratchet DB and the return of premium DB. The
ratchet DB has the feature that on each ratchet date, the death benefit
base will increase to the current account value, provided the account
value is higher. Let

\[
0\leq t_{1}<t_{2}<\dots<t_{m}\leq T\] 

\noindent
represent the set of ratchet dates prior to maturity. Then the rescaled
set, in terms of binomial time periods, is

\[
I=\left\{ \frac{t_{1}}{\delta t},\frac{t_{2}}{\delta t},\dots,\frac{t_{m}}{\delta t}\right\} \subset\mathcal{I}_{N}.\] 

\noindent

The GMWB and GMDB are treated as one rider with the aim of solving
for the fair fee $\alpha^{\star}$ as before. Alternatively, one could
separate the two and specify the GMDB rider fee exogenously. Let $DB_{i}$
be the death benefit guarantee base at time point $i$, with $DB_{0}=P$.
Then $DB_{i}=db(i,W_{i^{-}},DB_{i-1})$, where $db:\mathcal{I}_{N}\times\mathbb{R}_{+}\times\mathbb{R}_{+}\mapsto\mathbb{R}_{+}$
is defined as 

\begin{equation}
\begin{cases}
db(0,x,y)=x,\\
db(i,x,y)=\max\left(w(x)\mathbf{1}_{\{i\in I\}},\frac{w(x)}{xe^{-\alpha}}y\right).\end{cases}\label{eq:DB}\end{equation}

\noindent
If $I=\emptyset$, then the ratchet DB reduces to a simple return
of premium DB.

Note that $DB_{i}=0$ for $i\geq\tau$. However we assume that conditional
on survival to the trigger date, the guaranteed payments are paid
regardless of life status; that is, the present value of the remaining
payments is paid upon death if trigger has previously occurred. The
death benefit of $\max(DB_{i},W_{i+1^{-}})$ is paid at time $(i+1)\delta t$,
if death occurs during the $(i+1)$th period but prior to trigger
time. In the limit as $\delta t\to0$ this corresponds to the death
benefit being paid at the instantaneous time of death.

The death benefit base in \prettyref{eq:DB} is reduced by withdrawals
in a pro-rata manner, meaning it is reduced by the same proportion
as the account value. Another method is called dollar-for-dollar withdrawal
adjusted. Assume a policyholder holds a deep in the money GMDB, with
$DB_{i}\gg W_{i}$ (where $x\gg y$ means $y$ is much less than $x$).
By withdrawing $0.9W_{i}$ and ignoring surrender charges, under the
dollar-for-dollar reduction method the policyholder holds a GMDB with
only 10\% of the previous account value but a death benefit base of
$DB_{i}-0.9W_{i}\gg0$. Under the pro-rata method, the new death benefit
base is $0.1DB_{i}\ll DB_{i}-0.9W_{i}$.

\subsection{\label{sec:ValuationMortNo-Lapses-Model}Pricing and Hedging a Single Contract}

A key underlying assumption for the remainder of our work is stated. 
\begin{Assumption}
\label{ass:productspace}There is independence between biometric and
financial risks. Let $(\Omega^{S},\mathcal{F}_{N}^{S},\mathbb{F}^{S},\mathbb{Q}^{S})$
and $(\Omega^{M},\mathcal{F}_{N}^{M,\{x,l_{x}\}},\mathbb{F}^{M,x},\mathbb{P}^{M})$
be the filtered probability spaces constructed in \prettyref{sec:Bin wo mort}
and \prettyref{sec:Mortality-Risk-Model} respectively. We work with
the product space $(\Omega,\mathcal{F}_{N},\mathbb{F},\mathbb{Q})$
where $\Omega:=\Omega^{M}\times\Omega^{S}$, $\mathbb{F}:=\{\mathcal{F}_{i}\}_{i=0}^{N}$,
$\mathcal{F}_{i}:=\mathcal{F}_{i}^{M,\{x,l_{x}\}}\times\mathcal{F}_{i}^{S}:=\sigma(\{A\times B:\ A\in\mathcal{F}_{i}^{M,\{x,l_{x}\}},\ B\in\mathcal{F}_{i}^{S}\})$
and $\mathbb{Q}:=\mathbb{P}^{M}\times\mathbb{Q}^{S}$.
\end{Assumption}
We present the more general model allowing for early surrenders and
as in \prettyref{sec:Binlapses} optimal policyholder behaviour is
assumed. The no-lapse model is obtained under the following assumption.
\begin{Assumption}
\label{ass:AdmLaps}(No-lapse model) The surrender charges satisfy
$k_{i}=1$ for all $i<N$ and $k_{N}=0$. This implies that the set
of admissible lapse strategies is $\mathbb{L}_{0}=\{N\}$.
\end{Assumption}
Without loss of generality, from now until after \prettyref{sub:mort-divers}
we consider the case of a single contract sold to an individual at age $x$, that is we let $l_{x}=1$. The value process $\{V_{i}^{M}\}_{0\leq i\leq N}$
is defined as

\begin{align*}
  V_{i}^{M} & =A_{i}^{x}\max_{\eta\in\mathbb{L}_{i,\bar{\tau}_{i}}}E_{\Q}\Big[D_{\bar{\tau}_{i}\wedge\eta}^{x}\left(\max(DB_{K^{x}-1},W_{K^{x-}})e^{-\barr(K^{x}-i)}+Ga_{\lcroof{K^{x}-1-i}}\right)
    \\& \quad
    +A_{\bar{\tau}_{i}\wedge\eta}^{x}\left(Ga_{\lcroof{\eta-i}}+W_{\eta}(1-k_{\eta})e^{-\barr(\eta-i)}\right)|\mathcal{F}_{i}\Big].\end{align*}

\noindent
Observe that all $\eta\in\mathbb{L}_{i}$ are $\mathbb{F}^{S}$-stopping
times and are independent of the mortality probability measure. Any
lapse strategy $\eta$ is only exercised if the insured is still alive.
It remains true that the optimal lapse strategy must lie in $\mathbb{L}_{i,\bar{\tau}_{i}}\subset\mathbb{L}_{i}$.

Conditioning on the time of death and taking the expectation with respect to
$\mathbb{P}^{M}$ (justified by the independence of $\mathbb{Q}^{S}$
and $\mathbb{P}^{M}$) we obtain

\[
V_{i}^{M}=A_{i}^{x}V_{i},\] 

\noindent
where

\[
V_{i}=\max_{\eta\in\mathbb{L}_{i,\bar{\tau}_{i}}}V_{i}^{\eta}\] 

\noindent
and

\begin{align}
V_{i}^{\eta} & =E_{\mathbb{Q}^{S}}\Bigg[\sum_{j=i}^{\bar{\tau}_{i}\wedge\eta-1}{}_{j-i|}q_{x+i}\left(\max\left(DB_{j},W_{j+1^{-}}\right)e^{-\barr(j+1-i)}+Ga_{\lcroof{j-i}}\right)\label{eq:VoMortlapse}\\
 & \quad+{}_{\bar{\tau}_{i}\wedge\eta-i}p_{x+i}\left(Ga_{\lcroof{\eta-i}}+W_{\eta}(1-k_{\eta})e^{-\barr(\eta-i)}\right)\Big|\mathcal{F}_{i}^{S}\Bigg].\nonumber \end{align}

\noindent

The definition for the fair fee rate $\faira$ remains unchanged and
it satisfies $V_{0}^{M}=P$. Select any $\eta\in\mathbb{L}_{0}$.
Denote $^{R}\tilde{V}_{i}^{\eta}$ to be the total contract payouts
up to time point $i$ under this surrender strategy and discounted
to $t=0$. Then

\[
^{R}\tilde{V}_{i}^{\eta}=\sum_{j=0}^{\tau\wedge\eta\wedge i-1}\left(A_{j}^{x}-A_{j+1}^{x}\right)\left[\max(DB_{j},W_{j+1^{-}})e^{-\barr(j+1)}+Ga_{\lcroof{j}}\right]+A_{\tau\wedge\eta\wedge i}^{x}Ga_{\lcroof{\eta\wedge i}}.\] 

\noindent
Let $^{R}V_{i}^{\eta}:=E_{\mathbb{P}^{M}}[{}^{R}\tilde{V}_{i}^{\eta}]$.
Then we have

\[
^{R}V_{i}^{\eta}=\sum_{j=0}^{\tau\wedge\eta\wedge i-1}{}_{j|}q_{x}\left[\max(DB_{j},W_{j+1^{-}})e^{-\barr(j+1)}+Ga_{\lcroof{j}}\right]+{}_{\tau\wedge\eta\wedge i}p_{x}Ga_{\lcroof{\eta\wedge i}}.\] 

\noindent
For any $0\leq i\leq N,$ define a rescaled filtration $\mathbb{F}^{S,i}=\{\mathcal{F}_{j}^{S,i}:=\mathcal{F}_{j+i}^{S};0\leq j\leq N-i\}$.
Then the process

\begin{equation}
Y^{\eta,i}=\left\{ Y_{j}^{\eta,i}=e^{-\barr((j+i)\wedge\eta)}{}_{(j+i)\wedge\eta}p_{x}V_{(j+i)\wedge\eta}^{\eta}+{}^{R}V_{(j+i)\wedge\eta}^{\eta}\right\} _{0\leq j\leq N-i}\label{eq:mtgale-1}\end{equation}

\noindent
is a $(\mathbb{Q}^{S},\mathbb{F}^{S,i})$ martingale. The optimal
surrender strategy, $\hat{\eta}_{i}$, is given by \prettyref{eq:etahat}
(the proof is similar and uses the martingale \prettyref{eq:mtgale-1}). 

Since $\{W_{i},DB_{i}\}_{i=0,1,\dots N}$ is a 2-dimensional Markov
process we have

\[
V_{i}^{M}=A_{i}^{x}v(i,W_{i},DB_{i}),\] 

\noindent
where $v:\mathcal{I}_{N}\times\mathbb{R}_{+}\times\mathbb{R}_{+}\mapsto\mathbb{R}_{+}$
is recursively defined by

\[
v(N,x,y)=x\] 

\noindent
 and for $0\leq i\leq N-1$ 

\begin{alignat*}{1}
v(i,x,y) & =\max\{e^{-\barr}[p_{x+i}(G+pv(i+1,w(ux),db(i+1,ux,y))\\
 & \quad+qv(i+1,w(dx),db(i+1,dx,y)))\\
 & \quad+q_{x+i}((p\max(y,ux)+q\max(y,dx))\mathbf{1}_{\{x>0\}}+\mathbf{1}_{\{x=0\}}Ga_{\lcroof{N-i}})],x(1-k_{i})\}.\end{alignat*}

\noindent
This implies the boundary condition $v(i,0,y)=Ga_{\lcroof{N-i}}$.

The rider value process must account for the following cash flow components.
The rider fee is paid prior to trigger while the insured is alive
and has not surrendered. If surrender occurs prior to trigger time
then no cost is incurred for the GMWB rider. In the event that no
surrender occurs and the insured is alive at trigger time, the periodic
GMWB guarantee is paid out until maturity regardless of death. If
death occurs prior to the earlier of trigger time or surrender time,
then any excess of the death benefit over the current account value
is a cost incurred by the rider. Putting this together, we have

\begin{align}
U_{i}^{M} & =A_{i}^{x}\max_{\eta\in\mathbb{L}_{i,\bar{\tau}_{i}}}E_{\mathbb{Q}}\bigg[\sum_{j=i+1}^{\eta}e^{-\barr(j-i)}\Big[A_{\bar{\tau}_{i}}^{x}\left(G-W_{j^{-}}e^{-\bara}\right)^{+}-A_{j}^{x}W_{j^{-}}\left(1-e^{-\bara}\right)\nonumber \\
 & \quad-k_{\eta}W_{\eta}e^{-\barr(\eta-i)}A_{\eta}^{x}\Big]+D_{\eta}^{x}\left(DB_{K^{x}-1}-W_{K^{x-}}\right)^{+}e^{-\barr(K^{x}-i)}|\mathcal{F}_{i}\bigg].\label{eq:Umortlapse}\end{align}

\noindent
Then $U_{i}^{M}=A_{i}^{x}U_{i}=A_{i}^{x}u(i,W_{i},DB_{i})$, where
$u:\mathcal{I}_{N}\times\mathbb{R}_{+}\times\mathbb{R}_{+}\mapsto\mathbb{R}$
is described by

\begin{equation}
\begin{cases}
u(N,x,y)=0,\\
u(i,x,y)=\max\{e^{-\barr}(pu^{-}(i+1,ux,y)+qu^{-}(i+1,dx,y)),-k_{i}x\},\end{cases}\label{eq:umortlapse}\end{equation}

\noindent
and $u^{-}:\mathcal{I}_{N}^{+}\times\mathbb{R}_{+}\times\mathbb{R}_{+}\mapsto\mathbb{R}$
is given by

\begin{align}
u^{-}(i,0,y) & =G\ddot{a}_{\lcroof{N-i+1}},\nonumber \\
u^{-}(i,x,y) & =p_{x+i-1}[(G-xe^{-\bara})^{+}-x(1-e^{-\bara})+u(i,w(x),db(i,x,y))]\label{eq:uminusmortlapse}\\
 & \quad+q_{x+i-1}(y-x)^{+}.\nonumber \end{align}

\noindent
The notation $\ddot{a}_{\lcroof{i+1}}=1+a_{\lcroof{i}}$ is an annuity
due. Under \prettyref{ass:AdmLaps} it is easy to check that the term
$(-k_{i}x)$ is never binding. Note that $A_{i-1}^{x}u^{-}(i,W_{i^{-}},DB_{i-1})$
is $\mathcal{F}_{i}^{S}\times\mathcal{F}_{i-1}^{M,\{x,l_{x}\}}$-measurable.
It is the rider value at time point $i$ evaluated once the market
movement for the past period is known, but prior to any transactions
occurring (i.e.\ fees, withdrawals or death benefits). That is, the
insurer knows the exact market growth in the funds over the past period
but is waiting to find out about the status of the policyholder.

We denote $\{U_{i}^{M,NL}\}$ to refer to \prettyref{eq:Umortlapse}
when \prettyref{ass:AdmLaps} is in place. The marginal rider value
from the option to surrender is $L_{i}^{M}:=U_{i}^{M}-U_{i}^{M,NL}\geq0$
and can be written as

\begin{align}
L_{i}^{M} & =A_{i}^{x}\max_{\eta\in\mathbb{L}_{i,\bar{\tau}_{i}}}E_{\mathbb{Q}}\bigg[\sum_{j=\eta+1}^{N}e^{-\barr(j-i)}\left[A_{j}^{x}W_{j^{-}}\left(1-e^{-\bara}\right)-A_{\bar{\tau}_{i}}^{x}\left(G-W_{j^{-}}e^{-\bara}\right)^{+}\right]\nonumber \\
 & \quad-A_{\eta}^{x}\left[k_{\eta}W_{\eta}e^{-\barr(\eta-i)}+D_{N}^{x}\left(DB_{K^{x}-1}-W_{K^{x-}}\right)^{+}e^{-\barr(K^{x}-i)}\right]|\mathcal{F}_{i}\bigg].\label{eq:Umortlapse-1}\end{align}

\noindent
Then $L_{i}^{M}=A_{i}^{x}l(i,W_{i},DB_{i})$, where $l:\mathcal{I}_{N}\times\mathbb{R}_{+}\times\mathbb{R}_{+}\mapsto\mathbb{R}_{+}$
is given by

\begin{align*}
l(N,x,y) & =0,\\
l(i,x,y) & =\max\{p_{x+i}e^{-\bar{r}}(pl(i+1,w(ux),db(i+1,ux,y))\\
 & \quad+ql(i+1,w(dx),db(i+1,dx,y))),-u^{NL}(i,x,y)-k_{i}x\}.\end{align*}

\noindent
Backward induction verifies that $l(i,x,y)=u(i,x,y)-u^{NL}(i,x,y)$. 
\begin{Proposition}
\label{pro:VUWMortLapse}For any $\alpha>0$ we have 

\begin{equation}
V_{i}^{M}=U_{i}^{M}+A_{i}^{x}W_{i}\label{eq:VUWMortLaps}\end{equation}

\noindent
or equivalently 

\begin{equation}
V_{i}^{M}=U_{i}^{M,NL}+L_{i}^{M}+A_{i}^{x}W_{i}\label{eq:VULWMortLaps}\end{equation}

\noindent
$\mathbb{Q}$-a.s. for all $0\leq i\leq N$.\end{Proposition}
\begin{proof}
The equality \prettyref{eq:VUWMortLaps} can be proved either directly
from \prettyref{eq:VoMortlapse} and \prettyref{eq:Umortlapse} or
through backward induction applied to the functions $v$, $u$, and
$u^{-}$. The procedure is similar to the proof of \prettyref{thm:V=00003DU_WBin}.
We omit the details. 
\end{proof}
The $\mathbb{F}^{s}$-adapted portfolio process $\{\Delta_{i}\}$
is defined by $\Delta_{i}=\Delta(i,S_{i},W_{i},DB_{i})$, where $\Delta:\mathcal{I}_{N-1}\times\mathbb{R}_{+}^{3}\mapsto\mathbb{R}$
is given by

\begin{align}
\Delta(i,w,x,y) & =\frac{u^{-}(i+1,ux,y)-u^{-}(i+1,dx,y)}{wu-wd}.\label{eq:DeltaMortLapse}\end{align}

\noindent
Note that $\Delta(i,w,0,y)=0$. For a given policy, the insurer follows
$\{\Delta_{i}\}$ only up until the death of the policyholder or the
surrender of the policy.

Similar to \prettyref{sec:Binlapses}, we define a consumption process
$\{C_{i}\}_{0\leq i\leq N-1}$ where $C_{i}=c(i,W_{i},DB_{i})$ and
$c:\mathcal{I}_{N}\times\mathbb{R}_{+}\times\mathbb{R}_{+}\mapsto\mathbb{R}_{+}$
is defined as

\begin{align}
c(i,x,y): & =v(i,x,y)-e^{-\barr}[p_{x+i}(G+pv(i+1,w(ux),db(i+1,ux,y))\nonumber \\
 & \quad+qv(i+1,w(dx),db(i+1,dx,y)))]\nonumber \\
 & \quad+q_{x+i}((p\max(y,ux)+q\max(y,dx))\mathbf{1}_{\{x>0\}}+\mathbf{1}_{\{x=0\}}Ga_{\lcroof{N-i}})]\nonumber \\
 & =u(i,x,y)-e^{-\barr}[pu^{-}(i+1,ux,y)+qu^{-}(i+1,dx,y)].\label{eq:cmortu}\end{align}

\noindent
The second equality can be verified using \prettyref{pro:VUWMortLapse},
similar to \prettyref{eq:Consumption}. Under \prettyref{ass:AdmLaps}
we have $C\equiv0$. 

Construct the replicating portfolio by starting with initial capital
$X_{0}=x_{0}$ and following the portfolio process $\{\Delta_{i}\}$.
For $i\in\mathcal{I}_{N}^{+}$ we have

\begin{multline}
X_{i}=\left(X_{i-1}-A_{i-1}^{x}(\Delta_{i-1}S_{i-1}+C_{i-1})\right)e^{\barr}+A_{i-1}^{x}\Delta_{i-1}S_{i}+A_{i}^{x}\left[F_{i}-\left(G-W_{i^{-}}e^{-\bara}\right)^{+}\right]\\
-(A_{i-1}^{x}-A_{i}^{x})\left[(DB_{i-1}-W_{i^{-}})^{+}\mathbf{1}_{\{\tau\geq i\}}+G\ddot{a}_{\lcroof{N-i+1}}\mathbf{1}_{\{\tau<i\}}\right].\label{eq:Xmort}
\end{multline}

\noindent
The fees, payouts, portfolio process, and consumption process have
all been defined in $\mathbb{F}^{S}$. Of course they are only applicable
while the policy is in force (prior to death or surrender). For that
reason, the terms are accompanied by $A_{i}^{x}$ factors in \prettyref{eq:Xmort}.
Given a surrender strategy $\eta\in\mathbb{L}_{0},$ the insurer will
close out its position at time point $\eta$ and the process of interest
is $\{X_{i\wedge\eta}\}_{0\leq i\leq N}$. The time zero profit is
$\Pi=e^{-\barr\eta}X_{\eta}$, since if death occurs prior to $\eta$
then the portfolio remains unchanged for all periods between death
and $\eta$, aside from interest accumulation. 

Although we no longer have almost sure equivalence of $U^{M}$ and
$X$ with respect to the product measure $\mathbb{Q}$, an analogous
result holds by considering the conditional expectation with respect
to $\mathbb{P}^{M}$. 
\begin{Theorem}
\label{thm:limitmort}Suppose the fee rate $\alpha$ is charged and
the initial capital is $x_{0}=U_{0}^{M}$. Then the following relation
holds between $X_{i}$, described by \prettyref{eq:Xmort}, and $U_{i}^{M}$,
given by \prettyref{eq:Umortlapse}:

\[
\mathbb{Q}^{S}(\mathbb{E}_{\mathbb{P}^{M}}[X_{i}-U_{i}^{M}]=0)=1\] 

\noindent
 for all $i\in\mathcal{I}_{N}$. \end{Theorem}
\begin{proof}
See Appendix~\ref{Proofs}
\end{proof}

\subsection{\label{sub:mort-divers} Diversification of Mortality Risk for Multiple Contracts}

Suppose homogeneous policies are sold to $l_{x}$ independent policyholders
aged $x$, each with an initial premium of $P$ and the fair rider
fee $\alpha^{\star}$ is charged. For the pool of $l_{x}$ insureds,
the number of deaths between time $i\delta t$ and $(i+1)\delta t$
is

\[
\mathcal{D}_{i}^{l_{x},x}:=\sum_{j=1}^{l_{x}}\left(A_{i}^{x,j}-A_{i+1}^{x,j}\right)\] 

\noindent
for $i\in\mathcal{I}_{N-1}$. The number of members alive at time
$i$ is

\[
\mathcal{A}_{i}^{l_{x},x}=\sum_{j=1}^{l_{x}}A_{i}^{x,j}=l_{x}-\sum_{j=1}^{i-1}\mathcal{D}_{j}^{l_{x},x}.\] 

\noindent
By the strong law of large numbers (SLLN), as $l_{x}\to\infty,$

\[
\frac{\mathcal{D}_{i}^{l_{x},x}}{l_{x}}\to{}_{i|}q_{x}\quad\text{and}\quad\frac{\mathcal{A}_{i}^{l_{x},x}}{l_{x}}\to{}_{i}p_{x}\] 

\noindent
$\mathbb{P}^{M}$-a.s., for all $i\in\mathcal{I}_{N}$.

The aggregate replicating portfolio process is the sum of the individual
replicating portfolio processes given by \prettyref{eq:Xmort}:

\[
  X_{i}^{\{l_{x}\}}=\sum_{j=1}^{l_{x}}X_{i}^{j},
\] 

\noindent
where $X_{i}^{j}\in\mathcal{F}_{i}^{S}\times\mathcal{F}_{i}^{M,x,j}$
for $1\leq j\leq l_{x}$ and $1\leq i\leq N$. The aggregate rider
value process is

\[
U_{i}^{M,\{l_{x}\}}=\sum_{j=1}^{l_{x}}U_{i}^{M,j}=\mathcal{A}_{i}^{l_{x},x}U_{i},
\] 

\noindent
since $U_{i}^{j}=0$ if $A_{i}^{x,j}=0$. 
We define two processes
$\{X_{i}^{\star}=E_{\mathbb{P}^{M}}[X_{i}^{\{1\}}]\}_{i=0}^{N}$
and 
$\{U_{i}^{\star}=E_{\mathbb{P}^{M}}[U_{i}^{M,\{1\}}]\}_{i=0}^{N}$,
both of which lie in $(\Omega^{S},\mathcal{F}_{N}^{S},\mathbb{F}^{S},\mathbb{Q}^{S})$.
Then by the SLLN we have 

\begin{gather*}
\left\{ \frac{X_{i}^{\{l_{x}\}}}{l_{x}}\right\} \to\{X_{i}^{\star}\}\quad\text{and}\quad\left\{ \frac{U_{i}^{M,\{l_{x}\}}}{l_{x}}\right\} \to\{U_{i}^{\star}\}
\end{gather*}

\noindent
$\mathbb{P}^{M}$-a.s., as $l_{x}\to\infty$. Beginning with $X_{0}^{\star}=0$,
from \prettyref{eq:xstarmort} we have

\begin{align*}
X_{i}^{\star} & =X_{i-1}^{\star}e^{\barr}+{}_{i-1}p_{x}\Big[\Delta_{i-1}(S_{i}-S_{i-1}e^{\barr})-C_{i-1}e^{\barr}+p_{x+i-1}\left[F_{i}-\left(G-W_{i^{-}}e^{-\bara}\right)^{+}\right]\\
 & \quad-q_{x+i-1}\left[(DB_{i-1}-W_{i^{-}})^{+}\mathbf{1}_{\{\tau\geq i\}}+G\ddot{a}_{\lcroof{N-i+1}}\mathbf{1}_{\{\tau<i\}}\right]\Big]\end{align*}

\noindent
for $i\in\mathcal{I}_{N}^{+}$. It is immediate that $U_{i}^{\star}={}_{i}p_{x}U_{i}.$
Finally, from \prettyref{thm:limitmort} we have

\[
X_{i}^{\star}=U_{i}^{\star}\] 

\noindent
$\mathbb{Q}^{S}$-a.s., for $i\in\mathcal{I}_{N}$.

Mortality risk diversification is attained in the limit as $l_{x}\to\infty$,
and we have perfect hedging. The fair fee was determined assuming
optimal surrender behaviour on the part of each policyholder, given
survival. If policyholders act irrationally then the insurer can consume
from each portfolio at each occurrence of this irrationality. The
limiting aggregate portfolio process for the pool is constructed on
the basis of homogeneous behaviour of all policyholders, whether or
not they act rationally. 
\begin{Remark}
The limiting process was obtained assuming homogeneous policies. This
assumption can be weakened to allow for varying initial premiums $P$
by policy, although each policy must have an issue age of $x$ and
a common rider fee $\alpha$. This is true since $P$ can be scaled
out of all the processes and the rider fee is independent of the premium
$P$. Let the premium for policy $i$ be $P_{i}$. Suppose $\{P_{i};i\geq1\}$
satisfies $\sum_{i=1}^{n}P_{i}\to\infty$ as $n\to\infty$. Further
assume that $\{P_{i}\}$ is monotonically increasing and satisfies
$\sup_{n\geq1}\frac{nP_{n}}{\sum_{i=1}^{n}P_{i}}<\infty$ or that
$\{P_{i}\}$ is monotonically decreasing in which case no condition
is needed. From Theorem 1 in \citet{RefWorks:66}, as $l_{x}\to\infty$,
we have

\[
\frac{\sum_{j=1}^{l_{x}}P_{j}A_{i}^{x,j}}{\sum_{j=1}^{l_{x}}P_{j}}\to{}_{i}p_{x}\] 

\noindent
$\mathbb{P}^{M}$-a.s. for all $i\in\mathcal{I}_{N}$. Therefore

\[
\Bigg\{\frac{X_{i}^{\{l_{x}\}}}{\sum_{i=1}^{l_{x}}P_{i}}\Bigg\}\to\{X_{i}^{\star}\},\] 

\noindent
with a similar result for $U^{\star}$. The average is taken on a
per premium dollar basis and both $X^{\star}$and $U^{\star}$ have
$P=1$.
\end{Remark}

\subsection{Numerical Results}

We consider two examples. The mortality is modelled using \prettyref{exa:The-parameters-used}.
\begin{Example}
\prettyref{fig:Fees vs age} plots the fair rider fee $\alpha^{\star}$
against the issue age $x$ for a GMWB with a return of premium DB
and an annual ratchet DB without lapses. The parameters are: $g=7.14\%,\ T=14,\ r=5\%,\ \sigma=20\%$,
and $\delta t=1$. The ratchet adds considerably more value to the
contract. The figure on the right zooms in on the ages 40-70. The
GMWB plus return of premium DB rider is largely insensitive to $x$.
The payouts upon death or survival are fairly similar in this instance.
Under the binomial model  without
mortality, we have $\faira=53$bps or $V_{0}(100,53\text{bps})=100$.
For the return of premium DB with $x=60$, we have $\faira=58$bps
and $V_{0}^{M}(100,53\text{bps})=100.35$. Depending on the product
specifications and parameters, mortality may have only a small effect.%
\begin{figure}
\begin{centering}
\includegraphics{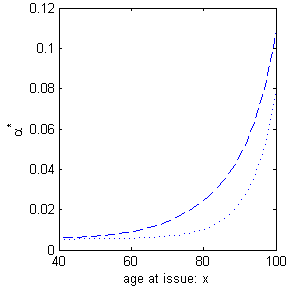}\includegraphics{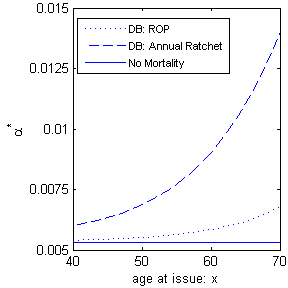}
\par\end{centering}

\caption[Fair rider fee as a function of issue age]{\label{fig:Fees vs age} $\faira$ as a function of issue age $x$}

\end{figure}

\end{Example}

\begin{Example}
The diversifiable mortality risk assumption is often quick to be used
in the literature. Given the prescribed portfolio process \prettyref{eq:DeltaMortLapse}
which assumes the risk is diversifiable, we consider the hedging losses
when there are only a finite number of policies sold. For $l_{x}\in\{10,1000,100000\}$
we simulated the time of deaths for each policy to obtain $\{\widehat{T}_{j}^{x}\}_{1\leq j\leq l_{x}}$,
and computed the average losses per policy per \$100 premium for each
path in the binomial model. The parameters used are: $x=60,$ $g=10\%,\ T=10,\ r=5\%,$
$\sigma=15\%$, $\delta t=1$, and $P=100$. Surrenders are not allowed.

For the GMWB with an annual ratchet DB, \prettyref{fig:Convergence-of-losses}
shows the convergence of the hedging losses to zero under the delta
hedging strategy as $l_{x}$ increases. The values are time-zero present
values and the losses under no hedging are also displayed. \prettyref{fig:Limiting-distribution-of}
plots the losses for the limiting portfolio $X^{\star}$. \prettyref{tab:MortHedgeStat}
provides the profit metrics $E_{\Q}[\Pi|\{\widehat{T}_{j}^{x}\}_{1\leq j\leq l_{x}}]$
and $SD_{\Q}[\Pi|\{\widehat{T}_{j}^{x}\}_{1\leq j\leq l_{x}}]$ for
hedging and no hedging where $\Pi$ is the average profit per policy
discounted to $t=0$. The results are also given when the DB rider
is a return of premium (ROP). The results for both DBs were obtained using the same sets
of simulated death times. The column with $l_{x}=\infty$ represents
the results for $X^{\star}$. The fair fee with the ratchet is 57bps
and with the ROP is 44bps. The metrics were calculated using the exact
binomial distribution under $\Q$ for the financial risk and the simulated
deaths for the mortality risk. For the purpose of examining convergence
with respect to $l_{x}$, we assume no market price of risk (i.e.\ $\mu=r$).

Selling a limited number of policies or facing capacity constraints
does not impose a significant risk to the insurer in this case because
the payouts are similar upon death or survival and diversification
occurs rapidly. The average hedging profits are higher with the ROP,
but the profits (losses) have more volatility with the ratchet since
it pays higher benefits and has higher fees. Under $\Q$ the expected
profits are equal under hedging and no hedging. It is the variance
that is reduced by hedging. %

\begin{figure}
\scalebox{0.8}{\includegraphics{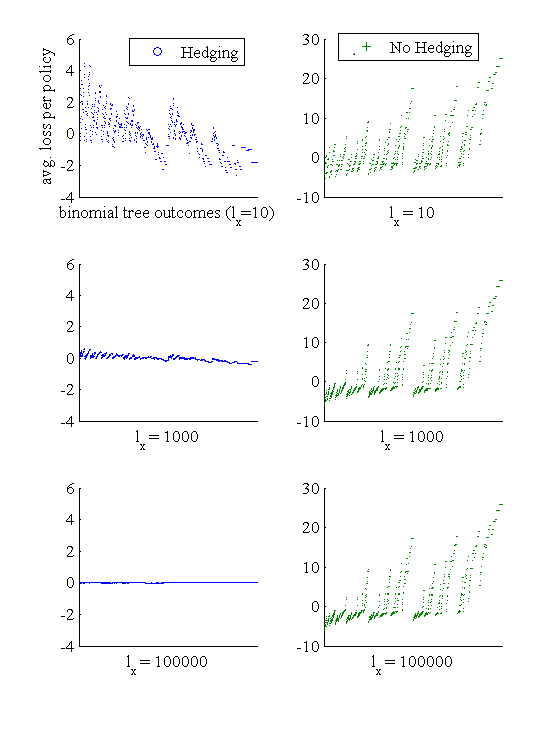}}
\caption[Convergence of losses for GMWB plus ratchet DB as $l_{x}\to\infty$]{\label{fig:Convergence-of-losses}Convergence of losses for GMWB
plus ratchet DB as $l_{x}\to\infty$ where the average losses per
policy under simulated mortality are shown for each market outcome.}

\end{figure}

\begin{figure}
\begin{centering}
\includegraphics{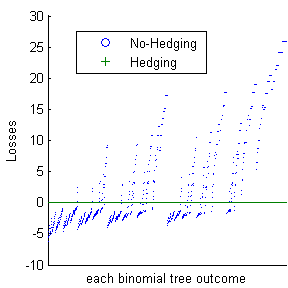}
\par\end{centering}
\caption{\label{fig:Limiting-distribution-of}Losses for GMWB plus ratchet
DB with complete diversification ($X^{\star})$}
\end{figure}

\begin{table}
\centering{}
\scalebox{0.9}{
\begin{tabular}{lccccccc} \toprule
Values per \$100 & \multicolumn{3}{c}{Hedging} & \multicolumn{4}{c}{No Hedging}\\\midrule
$l_x$ & 10 & 1000 & 100000 & 10 & 1000 & 100000 & $\infty$\\ \midrule
\multicolumn{8}{c}{GMWB + Ratchet DB}\\ \midrule
$E_{\Q} [\Pi|\{\widehat{T}_{j}^{x}\}_{1\leq j\leq l_{x}}]$ & 0.122 & 0.030 & 0.004 & 0.122 & 0.030 & 0.004 & 0\\
${SD}_{\Q} [\Pi|\{\widehat{T}_{j}^{x}\}_{1\leq j\leq l_{x}}]$ & 0.768  & 0.175& 0.008& 5.631& 5.787& 5.860 & 5.860\\ \midrule
\multicolumn{8}{c}{GMWB + Return of Premium DB}\\ \midrule
$E_{\Q} [\Pi|\{\widehat{T}_{j}^{x}\}_{1\leq j\leq l_{x}}]$ & 0.261 & 0.054 & 0.001 & 0.261 & 0.054 & 0.001 & 0\\
${SD}_{\Q} [\Pi|\{\widehat{T}_{j}^{x}\}_{1\leq j\leq l_{x}}]$ & 0.446  & 0.091& 0.004& 5.560& 5.736& 5.776 & 5.777\\ \bottomrule
\end{tabular}
}
\caption{\label{tab:MortHedgeStat}Profit metrics with and without hedging,
with GMDBs}

\end{table}

\end{Example}
Without mortality risk, each policy in the pool is subject to a common
equity risk and in the binomial world the correct hedging strategy
works for any number of policies. Mortality risk introduces incompleteness
into the model. Under the assumption of mortality risk diversification
the market regains completeness. This occurs in the limit by selling
sufficiently large pools of relatively small contract sizes.

Aside from risk pooling and diversification, other risk-management
options are reinsurance and longevity bonds. Additionally the typical
large life insurer with significant amounts of underwritten business
in life insurance and annuities has a degree of natural risk reduction
since these instruments have partially offsetting risks. Assuming
none of these option are available - there are no re-insurers, longevity
bonds do not exist and the insurer only sells annuities - the insurer's
main tool for mitigating its risk exposure is by selling a large number
of policies of relatively small amounts, thus reducing fluctuations
in the realized mortality rates around the expected mortality rates. 

\section{Conclusions}\label{sec:Conc}
In this paper we have constructed a binomial asset pricing model for the variable annuity with GMWB rider
which incorporated optimal policyholder surrender behaviour.  We extend the continuous time results of \citet{Hyndman-Wenger-Decomp} to the discrete-time binomial model by considering the valuation perspectives of the insured and the insurer.  These extensions allow us to prove the existence and uniqueness of the fair rider fee and decompose the value of the variable annuity with GMWB rider into term-certain payments and embedded derivatives.  Further, in the discrete time binomial model we are able to provide explicit perfect hedging strategies and optimal surrender strategies.

From a computational perspective the ability to model early surrenders using the basic tools of  binomial models is one distinct advantage over Monte Carlo methods. The other advantage was demonstrated by easily obtaining an explicit hedging strategy in a binomial (CRR) world that was proved to perfectly hedge the product. A drawback of the binomial model is the $O(2^N)$ growth of the non-recombining binomial trees. Nevertheless, by the tractability of the model and its finite nature, it is straightforward to obtain numerical results concerning any aspect of the product, provided that the number of time-steps is manageable. The qualitative conclusions drawn from such an analysis will usually hold true in the more general continuous model. We present comprehensive numerical results which are consistent with those presented in more complex models.

The binomial modeling framework is further extended to account for diversifiable mortality risk. The diversification argument for mortality risk is sometimes abused in the literature. After applying diversification arguments to obtain the fair fee and hedging results, we imposed capacity constraints by considering finite pools and saw that diversification occurs fairly rapidly. The results support the common claim that insurers are able to diversify mortality risk.

%% file: Hyndman-Wenger-Appendix-nc.tex
\section{\label{Proofs}Proofs of Technical Results}

\begin{proof}[Proof of \prettyref{lem:MonotBin}]
From the equivalent expression for $v(i,x)$ in \prettyref{eq:Vonovector},
the continuity result is immediate. The maximum possible value for
$W_{N}^{x,i}$ is obtained by the path corresponding to $\omega_{j}=u$
for all $j=i+1,\dots,N$. Thus 

\[
 b^{x,i}=\min\{\alpha\geq0:W_{N}^{x,i}(uu\dots u)=0\}.
\]

\noindent
From \prettyref{eq:W_Ndirect}, $W_{N}^{x,i}(uu\dots u)=0$ if and
only if 

\[
f(\alpha):=\left(x(e^{-\bara }u)^{N-i}-G\sum_{j=0}^{N-i-1}(e^{-\bara}u)^{j}\right)\leq0.\]

\noindent
But $f\in C^{\infty}$ and $\lim_{\alpha\to\infty}f(\alpha)=-G<0$.
We have $f(0)>0$ if and only if \prettyref{eq:x>Gsum...} holds.
If $f(0)>0$ then there exists $0<b^{x,i}<\infty$. If $f(0)\leq0,$
then $b^{x,i}=0$. The remainder of the proof is similar to the proof of
\citet[Lemma~4]{Hyndman-Wenger-Decomp}. Assume $(i,x)$ is such that $b^{x,i}>0$.
Let 

\[
A^{\alpha}:=\{W_{N}^{x,i}(\alpha)>0\}.\]

\noindent
Then $A^{\alpha}\neq\emptyset$ for $\alpha<b^{x,i}$. Fix $\alpha\in[0,b^{x,i})$
and consider $\alpha^{(1)}$ such that $\alpha<\alpha^{(1)}<b^{x,i}$.
When restricted to the set $A^{\alpha^{(1)}}$, \prettyref{eq:W_Ndirect}
implies 

\[
0<W_{N}^{x,i}(\alpha^{(1)})<W_{N}^{x,i}(\alpha),\]

\noindent
which in turn implies $A^{\alpha^{(1)}}\subseteq A^{\alpha}$. We
conclude that $v(i,x;\alpha^{(1)})<v(i,x;\alpha)$. 
\end{proof}
\begin{proof}[Proof of \prettyref{thm:existuniquealpha}]
From \prettyref{lem:MonotBin}, for $\alpha\geq b^{P,0}>0$, we have
$V_{0}(P,\alpha,g)=Ga_{\lcroof{N}}<P$ for $r>0$. By the definition
of $U$ in \prettyref{eq:Udiscrte} we have $U\geq0$ for $\alpha=0$.
By \prettyref{thm:V=00003DU_WBin}, 

\[
V_{0}(P,\alpha=0,g)=U_{0}(P,\alpha=0,g)+P\geq P.\]

\noindent
 By the continuity and strictly decreasing property from \prettyref{lem:MonotBin},
there exists a unique $\alpha^{\star}\in[0,b^{P,0})$. 
\end{proof}

\begin{proof}[Proof of \prettyref{thm:V=00003DU_WBin}]
We apply backward induction and show that $v(i,x)=u(i,x)+x$ for all
$(i,x)\in\mc{I}_{N}\times\mb{R}_{+}$. By definition $v(N,x)=u(N,x)+x$
for all $x\in\mb{R}_{+}$. Assume $v(i,x)=u(i,x)+x$ holds for
all $x\in\mb{R}_{+}$ for some $1\leq i\leq N$. We need to show
that $v(i-1,y)=u(i-1,y)+y$ for all $y\in\mb{R}_{+}$. Applying
the induction hypothesis,

\begin{align*}
v(i-1,y) & =e^{-\barr}[G+pv(i,w(uy))+qv(i,w(dy))]\\
 & =e^{-\barr}[pu(i,w(uy))+qu(i,w(dy))+p(w(uy)+G)+q(w(dy)+G)].
\end{align*}

\noindent

\noindent
From equations \prettyref{eq:UiShortVersion-1} and \prettyref{eq:u-} we have

\begin{align*}
&u(i-1,y) =e^{-\barr}\bigg\{pu(i,w(uy))+qu(i,w(dy))  \\ 
&+ p[(G-uye^{-\bara})^{+}-uy(1-e^{-\bara})]  +q[(G-dye^{-\bara})^{+}-dy(1-e^{-\bara})]\bigg\}. &
\end{align*}

\noindent
Observe 

\[
w(y)-(G-ye^{-\bara})^{+}=ye^{-\bara}-G.\]

\noindent
Then 

\[
w(y)+G-(G-ye^{-\bara})^{+}+y(1-e^{-\bara})=y,\]

\noindent
therefore

\begin{align*}
  v(i-1,y)-u(i-1,y) & =e^{-\barr}[puy+qdy]  =y
\end{align*}

\noindent
since $pu+qd=e^{\barr}$ by the definition of the risk-neutral probabilities
\prettyref{eq:RNP}.
Therefore

\[
v(i-1,y)=u(i-1,y)+y\]

\noindent
for all $y\in\mb{R}_{+}$ and the result holds.
\end{proof}

\begin{proof}[Proof of \prettyref{thm: Hedg GMWB-1}]
Following the approach of \citet{RefWorks:33}, we proceed
by induction. By assumption we have that $X_{0}=U_{0}$. Assume for
some $0\leq i<N$ that $X_{i}=U_{i}$. We need to show that for all
$\bar{\omega}_{i}$, 

\begin{align*}
X_{i+1}(\bar{\omega}_{i}u) & =U_{i+1}(\bar{\omega}_{i}u),\\
X_{i+1}(\bar{\omega}_{i}d) & =U_{i+1}(\bar{\omega}_{i}d).\end{align*}

\noindent
We omit the $\bar{\omega}_{i}$ notation for conciseness. Substituting
$U_{i}$ for $X_{i}$ in \prettyref{eq:Xilapse}, using \prettyref{eq:DeltaBinNolapse},
\prettyref{eq:Consumption}, and the fact $q=\frac{u-e^{\barr}}{u-d}$
we obtain

\begin{align*}
&X_{i+1}(u)  =\Delta_{i}S_{i}(u-e^{\barr})+(U_{i}-C_{i})e^{\barr}+F_{i+1}(u)-(G-W_{i+1^{-}}(u)e^{-\bara})^{+}\\
 & =q[u^{-}(i+1,uW_{i})-u^{-}(i+1,dW_{i})]+(pu^{-}(i+1,uW_{i})+qu^{-}(i+1,dW_{i})\\
 & \quad+F_{i+1}(u)-(G-W_{i}ue^{-\bara})^{+}\\
 & =u^{-}(i+1,uW_{i})+F_{i+1}(u)-(G-W_{i}ue^{-\bara})^{+}\\
 & =u(i+1,w(uW_{i}))\\
 & =U_{i+1}(u).\end{align*}

\noindent
A similar argument shows that $X_{i+1}(d)=U_{i+1}(d)$. Since $\bar{\omega}_{i}$
was arbitrary we have $X_{i+1}=U_{i+1}$ and the result holds.
\end{proof}

\begin{proof}[Proof of \prettyref{thm:limitmort}]
  We proceed by induction. By assumption we have that $X_{0}=U_{0}^{M}$.
Suppose that $E_{\mathbb{P}^{M}}[X_{i}]=E_{\mathbb{P}^{M}}[U_{i}^{M}]$
$\mathbb{Q}^{S}$-a.s.\ for some $i\in\mathcal{I}_{N-1}$. For a
process $H_{i}$ we write $H_{i}(\bar{\omega}_{i};j)$ for its value
at time $i$ for the specific path $\bar{\omega}_{i}\omega_{i+1}\dots\omega_{N}\in\Omega^{S}$
(where $\omega_{j}$ can take any value in $\{u,d\}$ for all $j>i$)
and the specific set $(K^{x})^{-1}(j)\in\mathcal{F}_{N}^{M,\{x,1\}}$.
For any fixed $\bar{\omega}_{i}$ we need to show that

\[
\begin{cases}
E_{\mathbb{P}^{M}}[X_{i+1}(\bar{\omega}_{i}u;K^{x})]=E_{\mathbb{P}^{M}}[U_{i+1}^{M}(\bar{\omega}_{i}u;K^{x})],\\
E_{\mathbb{P}^{M}}[X_{i+1}(\bar{\omega}_{i}d;K^{x})]=E_{\mathbb{P}^{M}}[U_{i+1}^{M}(\bar{\omega}_{i}d;K^{x})].\end{cases}\] 

\noindent
We prove the first equality, the second one is shown in an identical
manner. For conciseness, we omit $\bar{\omega}_{i}$. 

Observe that $E_{\mathbb{P}^{M}}[U_{i+1}^{M}(u;K^{x})]={}_{i+1}p_{x}U_{i+1}(u)$.
Also $X_{i+1}(u;j)=X_{i+1}(u;K^{x}>i+1)$ for all $j>i+1$, since
$X_{i+1}\in\mathcal{F}_{i+1}$. From \prettyref{eq:Xmort} we have
$X_{i+1}(u;j)=X_{i}(;j)e^{\barr}$ for all $j\leq i$. Therefore

\begin{align*}
E_{\mathbb{P}^{M}}[X_{i+1}(u,K^{x})] & =\sum_{j=1}^{N}{}_{j-1|}q_{x}X_{i+1}(u,j)+{}_{N}p_{x}X_{i+1}(u,K^{x}>N)\\
 & =\sum_{j=1}^{i}{}_{j-1|}q_{x}X_{i}(;j)e^{\barr}\negthinspace+\negthinspace{}_{i|}q_{x}X_{i+1}(u,i+1)\negthinspace+\negthinspace{}_{i+1}p_{x}X_{i+1}(u,K^{x}\negthinspace>\negthinspace i\negthinspace+\negthinspace1).\end{align*}

\noindent
After applying \prettyref{eq:Xmort} to $X_{i+1}(u,i+1)$ and $X_{i+1}(u;K^{x}>i+1)$,
we obtain

\begin{align}
E_{\mathbb{P}^{M}}[X_{i+1}(u,K^{x})] & =E_{\mathbb{P}^{M}}[X_{i}(;K^{x})]e^{\barr}+_{i}p_{x}[\Delta_{i}S_{i}(u-e^{\barr})-C_{i}e^{\barr}\nonumber \\
 & \quad-p_{x+i}((G-W_{i}ue^{-\bara})^{+}-W_{i}u(1-e^{-\bara}))\label{eq:xstarmort}\\
 & \quad-q_{x+i}((DB_{i}-W_{i}u)^{+}\mathbf{1}_{\{\tau>i\}}+G\ddot{a}_{\lcroof{N-i}}\mathbf{1}_{\{\tau\leq i\}})].\nonumber \end{align}

\noindent
By the induction hypothesis,

\[
E_{\mathbb{P}^{M}}[X_{i}(;K^{x})]e^{\barr}={}_{i}p_{x}U_{i}e^{\barr}.\] 

\noindent
 Then substituting \prettyref{eq:DeltaMortLapse} and \prettyref{eq:umortlapse}
and applying \prettyref{eq:uminusmortlapse} (in the form $U_{i+1}^{-}$,
but conditioning on $\tau>i$), we have

\begin{align*}
E_{\mathbb{P}^{M}}[X_{i+1}(u,K^{x})] & ={}_{i}p_{x}[(U_{i}-C_{i})e^{\barr}+(U_{i+1}^{-}(u)-U_{i+1}^{-}(d))q-Gp_{x+i}\mathbf{1}_{\{\tau\leq i\}}\\
 & \quad+\mathbf{1}_{\{\tau>i\}}(p_{x+i}U_{i+1}(u)-U_{i+1}^{-}(u))-q_{x+i}(G\ddot{a}_{\lcroof{N-i}}\mathbf{1}_{\{\tau\leq i\}})]\\
 & ={}_{i}p_{x}[U_{i+1}^{-}(u)\mathbf{1}_{\{\tau\leq i\}}-Gp_{x+i}\mathbf{1}_{\{\tau\leq i\}}+\mathbf{1}_{\{\tau>i\}}p_{x+i}U_{i+1}(u)\\
 & \quad-q_{x+i}G\ddot{a}_{\lcroof{N-i}}\mathbf{1}_{\{\tau\leq i\}}]\\
 & ={}_{i+1}p_{x}[\mathbf{1}_{\{\tau>i\}}U_{i+1}(u)+Ga_{\lcroof{N-(i+1)}}\mathbf{1}_{\{\tau\leq i\}}\\
 & ={}_{i+1}p_{x}U_{i+1}(u).\end{align*}

\noindent
This completes the proof.
\end{proof}